 \documentclass[twocolumn]{aastex6}

\usepackage{changepage} 
\usepackage{amsmath} 
\usepackage{soul}

\fullcollaborationName{The Friends of AASTeX Collaboration}

\begin{document}


\title{Dynamical analysis of the circumprimary planet \\ in the eccentric binary system HD~59686}


\author{Trifon Trifonov\altaffilmark{1,2},
Man Hoi Lee\altaffilmark{1,3}, Sabine Reffert\altaffilmark{4}, and Andreas Quirrenbach\altaffilmark{4}}

\affil{$^1$ Department of Earth Sciences, The University of Hong Kong, Pokfulam Road, Hong Kong}
\affil{$^2$ Max-Planck-Institut f\"{u}r Astronomie,\ K\"{o}nigstuhl  17, 69117 Heidelberg, Germany}
\affil{$^3$ Department of Physics, The University of Hong Kong, Pokfulman Road, Hong Kong}

\affil{$^4$ Landessternwarte, Zentrum f\"{u}r Astronomie der Universit\"{a}t Heidelberg, \ K\"{o}nigstuhl 12, 69117 Heidelberg, Germany}
\email{trifonov@mpia.de}



%
%
%
%




\begin{abstract}

We present a detailed orbital and stability analysis of the HD~59686 binary-star planet system. 
HD~59686 is a single-lined moderately close ($a_{B} = 13.6\,$AU) eccentric ($e_{B} = 0.73$) binary,
where the primary is an evolved K giant with mass $M = 1.9 M_{\odot}$ and 
the secondary is a star with a minimum mass of $m_{B} = 0.53 M_{\odot}$.
Additionally, on the basis of precise radial velocity (RV) data a Jovian
planet with a minimum mass of $m_p = 7 M_{\mathrm{Jup}}$, orbiting the primary on a
nearly circular S-type orbit with $e_p = 0.05$ and $a_p = 1.09\,$AU, has
recently been announced.
We investigate large sets of orbital fits consistent with HD 59686's radial velocity data
by applying bootstrap and systematic grid-search techniques coupled with self-consistent dynamical fitting. 
We perform long-term dynamical integrations of these fits to constrain the
permitted orbital configurations. We find that if the binary and the planet in this system have 
prograde and aligned coplanar orbits, there are narrow regions of stable orbital solutions
locked in a secular apsidal alignment with the angle between the periapses, $\Delta \omega$, librating about $0^\circ$.
We also test a large number of mutually inclined dynamical models in an attempt to 
constrain the three-dimensional orbital architecture.
We find that for nearly coplanar and retrograde orbits with mutual inclination $145^\circ \lesssim \Delta i \leq 180^\circ$,
the system is fully stable for a large range of orbital solutions. 

\end{abstract}




\section{Introduction}

The first Doppler surveys looking for extrasolar planets were focused on finding Solar
system analogs and usually avoided 
known binary stars with semi-major axes $a_B$~$\leq$~200~AU \citep[see][]{Eggenberger2010, Thebault2014}. 
As a result, the number of known binary systems with planets orbiting around
one of the components (in circumstellar or S-type orbits),
or orbiting around both stars \citep[in circumbinary or P-type orbits, see][]{Dvorak1986} 
is still relatively low when compared to planets orbiting single stars.
To date\footnote{http://www.univie.ac.at/adg/schwarz/multiple.html} 
we know of $\sim$ 50 S-type planets which are part of wide binaries 
separated by at least 50--1000 AU \citep{Roell2012} and $\sim$ 20 P-type planets orbiting both stars where the binary separation is below 1 AU
\citep[mostly discovered with the Kepler satellite,][]{Borucki2010, Doyle2011, Welsh2012, Leung2013,Kostov2013}.
However, only a handful of S-type planet candidates in moderately close binary systems ($a_B$~$\leq$~30~AU) 
are known in the literature and they all were discovered using the radial velocity (RV) method.

A famous example is the $\gamma$~Cephei binary system, which consists
of a K giant primary of $M$ = 1.6~$M_{\odot}$ and a secondary star with a minimum mass of $m_B \sin i$ = 0.44~$M_{\odot}$,
separated by $a_{B}$ $\sim$ 19 AU. This system has a Jovian planet with a minimum mass of
$m_p \sin i$ $\sim$ 1.7~$M_{\mathrm{Jup}}$ \citep{Campbell1988,Hatzes2003} 
orbiting on a stable orbit around the primary star at $a_p$ $\sim$ 2.0 AU \citep{Haghighipour2006}.
A planet candidate on an S-type orbit is also evident in the RV
data taken for the HD~196885 binary system \citep{Correia2008}. 
This system consists of an F8V primary of $M$ = 1.3~$M_{\odot}$ and
a secondary star with a minimum mass of $m_B \sin i$ = 0.45~$M_{\odot}$, orbital semi-major axis $a_{B}$ = 21 AU and eccentricity $e_B$ = 0.42.
The double Keplerian best-fit for the RV data of HD 196885 reveals an S-type planet around the primary 
with $a_p$ $\sim$ 2.6 AU, $e_p$ $\sim$ 0.48 and a minimum mass of $m_p \sin i$ $\sim$ 3.0~$M_{\mathrm{Jup}}$.
\citet{Chauvin2011} have carried out dynamical simulations which show that the planet's orbit is more stable
in a highly inclined configuration near the equilibrium points of the Lidov-Kozai regime
\citep{Lidov1962,Kozai1962}. Later, \citet{Giuppone2012} have confirmed the stability of a highly inclined
configuration with a mutual inclination of $\Delta i$ $\approx$ 43$^\circ$ or 137$^\circ$, 
but they also found stable nearly coplanar configurations, where the planet's orbit is either 
prograde or retrograde, with the retrograde orbits being less chaotic.

Another remarkable example is the $\nu$~Octantis binary \citep{Ramm2009,Ramm2015,Ramm2016}.
This system consists of a $M$ = 1.6~$M_{\odot}$ K1~III giant primary 
and a low-mass secondary star separated only by $a_B$ = 2.5 AU, with moderate eccentricity of $e_B$ = 0.24.
The binary inclination is well constrained at $i_B$~=~71$^\circ$, which yields a secondary mass of $m_B$ = 0.6~$M_{\odot}$.
A lower amplitude periodic RV variation is present in addition to the secondary star RV signal, and 
if these variations are due to an orbiting planet, then the S-type companion would have
$a_p \sim$ 1.2 AU, $e_p \sim$ 0.1 and a minimum mass of about $m_p \sin i$ $\sim$ 2.0~$M_{\mathrm{Jup}}$.
The planetary interpretation is problematic because the best-fit orbit (with
semi-major axis ratio $a_p$/$a_{B}$ $\approx$ 0.47) is located well outside the boundary for stability if
one assumes a coplanar and prograde planet with respect to the binary's orbit \citep{Holman1999}. 
However, \citet{Eberle2010} and \citet{Gozdziewski2013} have
shown that a nearly coplanar retrograde orbit is stable, even though the
stable region is small due to nearby mean-motion resonances (MMR) at the 2:1, 3:1, and 5:2 period ratios.

The existence of prograde, retrograde, or even Lidov-Kozai resonance S-type giant planets as a
part of moderately compact systems remains a very challenging dynamical problem. 
Apart from the long-term stability problem, it is puzzling how planets can grow through 
core-accretion or disk-instability mechanisms in such close binaries.
These systems provide important clues on how planets could form and remain in
stable orbits around a star under the strong gravitational influence of a close stellar companion.

In this work we study the HD~59686 single-lined binary system, which 
is composed of a 1.92~$M_{\odot}$ K-giant and a low-mass star with a minimum mass of $m_B \sin i$ = 0.53~$M_{\odot}$.
This system was reported to have a massive ($m_p \sin i$  $\approx$ 7.0~$M_{\mathrm{Jup}}$)
Jovian S-type planet orbiting at $a_p$~=~1.09~AU around the primary star \citep{Ortiz2016}.
The binary itself, however, is very eccentric ($e_B$ = 0.73), which challenges the planet's orbital stability. 
We carry out an extensive statistical and dynamical analysis to the available RV data to demonstrate that this 
system has stable configurations and to further constrain its orbital parameters.

This paper is organized as follows: In~Section~\ref{HIP36616} 
we review the physical configuration of the HD 59686 system.
Section~\ref{Dynamical fit} describes the methodology of our dynamical fitting and long-term stability analysis. 
In Section~\ref{best_fits} we introduce the best-fit results from our tests and we reveal the possible S-type planet configurations.
In Sections \ref{Statistical} and \ref{systematic} we present dynamical and stability results
around the best fits based on our bootstrap and systematic parameter grid search analysis.
Finally, in Section~\ref{Discussion} we present conclusions based on our results 
and discuss the possible S-type planet configurations.

\section{The HD~59686 binary-planet system}
\label{HIP36616}

\subsection{System configuration}

HD~59686 (= HR~2877, HIP 36616) is a bright, photometrically 
stable \citep[V~=~5.45 mag., ][]{Leeuwen} horizontal branch (HB)
red~giant star with an estimated mass of $M$~=~1.92~$\pm$~0.21~$M_{\odot}$, 
radius of $R$~=~13.2 $\pm$ 0.3~$R_{\odot}$ \citep{Reffert2014}
and metal abundance of [Fe/H]~=~0.15 $\pm$ 0.1 \citep{Hekker2}.
With luminosity $L$ = 73.3~$\pm$~3.3 $L_{\odot}$ and effective temperature $T_{\mathrm{eff}}$~=~4658~$\pm$~24~K, 
HD~59686 is a typical K2 III giant star. More physical parameters of HD~59686 can be found in \citet{Reffert2014}.

Based on 88 precise (5 -- 8 m\,s$^{-1}$) RV observations of HD~59686 taken at Lick Observatory between 
November 1999 and December 2011, \citet{Ortiz2016} reported that HD~59686 is actually part of a much more complicated three-body system. 
The Keplerian orbital solution for HD~59686 given in \citet{Ortiz2016} shows 
that the RV data has a large amplitude variation of $K_B$ = 4014.12~$\pm$~5.84~m\,s$^{-1}$, whose characteristic RV
shape reveals a stellar companion with $m_B \sin i$ = 0.53~$M_{\odot}$ on a highly eccentric orbit ($e_B$ = 0.73). 
The large eccentricity and the fact that the binary recently passed through its periastron ($\sim$ February 2008) 
allowed the orbital period to be well determined as $P_B$ = 11679.94~$\pm$~192.92 days, even though the 
time span of the observations does not cover a full period of the binary orbit.
In addition, the RV data yielded a lower-amplitude signal 
of $K_p$ = 136.92~$\pm$~3.31~m\,s$^{-1}$ with a derived period of $P_p$~=~299.36~$\pm$~0.28 days.
This signal is due to a Jovian planet with a minimum mass of $m_p \sin i$ = 6.92 $M_{\mathrm{Jup}}$ 
on a nearly circular ($e_p$~=~0.05~$\pm$~0.02) S-type orbit around the primary K~giant star. 
The planetary signal remained coherent over many periods and was further 
confirmed by \citet{Trifonov2015} using follow-up RV measurements in the near-infrared 
taken with ESO's VLT spectrograph CRIRES \citep{Kaeufl}.

\subsection{Constraints on the Planetary Companion from the Hipparcos
Intermediate Astrometric Data}
\label{Constrauints}

HD 59686 was a target of the Hipparcos mission (HIP 36616). We analyze the
Hipparcos Intermediate Astrometric Data of HD 59686 based on the re-reduction
by \citet{Leeuwen} in the same way as described in \citet{Reffert2011}.
We ignore the stellar companion, since its period is much longer than
the Hipparcos mission duration, and fit only the astrometric orbit of the
planetary companion to the abscissa residuals, simultaneously allowing for
adjustments in the standard astrometric parameters (position, proper motion,
parallax) and keeping the spectroscopic parameters fixed. The best fit occurs
at an inclination $i_p$ = 2.9$^\circ$ and longitude of the ascending
node $\Omega_p$ = 266.7$^\circ$. 
The joint 3$\sigma$ confidence region extends from $i_p$ = 1.6$^\circ$ to 12.8$^\circ$ 
and from $\Omega_p$ = 206.8$^\circ$ to 308.2$^\circ$. 
Thus, a significant parameter range in $i_p$ and $\Omega_p$ 
can formally be rejected as a possible solution, and in 
\citet{Reffert2011} we argued that this is the best indicator for an actual
detection of the astrometric orbit.

However, there are several concerns with the Hipparcos data of HD 59686.

(1) The reduced $\chi^2$ value in the van Leeuwen version of the Hipparcos Catalog
    is 0.71, already quite small. It indicates that either the errors are
    overestimated, or that the solution is in fact already quite satisfactory,
    with no need for a better model for the measurements. The reduced $\chi^2$
    value after fitting for the astrometric orbit is 0.59, which is
    uncomfortably small.

(2) There is a clear correlation between the time of the year and the scan
    direction (not only for HD 59686, but also for other targets). This is
    particularly a problem for periods close to one year, which is the case for
    the planet orbiting HD 59686 (best fit spectroscopic period $\approx$ 299
    days). On top of that, many Hipparcos measurements have been obtained
    during the same season, i.e.\ with the same scan direction. Thus, the
    Hipparcos measurements for HD 59686 are poorly constrained in the
    perpendicular direction (roughly coinciding with the right ascension
    direction), and can float freely to fit any astrometric orbit.

We believe that, as a result, the Hipparcos data for HD 59686 should be treated
with caution. In fact, we will show later on that any solution with
$i_p < 30^\circ$ for the inner companion is highly improbable, so the
detection of an astrometric orbit could not be brought in line with the
observed radial velocity data. We conclude that most likely the astrometric
orbit of the inner companion has not been detected in the Hipparcos data.

\subsection{Dynamical considerations}

As discussed in \citet{Ortiz2016}, the stellar companion of HD~59686 must be either 
a low-mass (and low-luminosity) star such as a K dwarf or a white dwarf remnant.
These scenarios are particularly important to trace the possible origin of the S-type planet.
However, the question of whether the secondary is 
a K dwarf or white dwarf is of little importance for the goal of this paper, which is 
to study the current permitted (stable) orbital configuration. 

At first look, it is unclear how the planet could remain stable in such configuration.
The binary semi-major axis is $a_B$~=~13.6~AU, but the pericenter distance is only $q_B$~=~3.67~AU, leading to strong
interactions with the planet, which has $a_p$~=~1.09~AU and $q_p$~=~1.03~AU. 
Assuming a minimum mass of $m_B \sin i$ = 0.53 $M_\odot$ the Hill radius of the secondary star can be approximated as:  

\begin{equation}
 r_{\rm H,B}\approx a_B\sqrt[3]{m_B / 3M_{\star}} \approx ~6.2\,{\rm AU},
\label{eq:hill1}
\end{equation}

\noindent
which would cover the S-type planet orbit entirely during the binary periastron passage.
Due to the large eccentricity of the binary, however, one can define the Hill radius 
at the pericenter distance $q_B$ instead of $a_B$ \citep[see][]{Hamilton1992}, which leads to a
smaller value of $r_{\rm H,B} \approx$ 1.66\, AU. 
This suggests that the planet-secondary separation close to the binary periastron would be $\sim$1.5\,$r_{\rm H,B}$,
making the survival of the planet still challenging.

A~quick check using the empirical stability criterion of \citet{Holman1999} 
reveals that the critical (upper limit) semi-major axis for the S-type planet is $a_{\mathrm{crit.}}$ $\sim$ 1.03 AU.  
Considering the binary-planet orbital uncertainties, we find that the S-type planet is most likely unstable,
with an orbit slightly outside the stability region. 
Using similar empirical stability criteria from \citet{Eggleton1995}, we find that the planet is most likely stable,
while the criterion of \citet{Mardling2001} suggests that the planet is unstable.
In any case, these stability criteria agree that the planet is close to the stability border.
Therefore, in this paper we aim to inspect the three-dimensional orbital architecture of the HD~59686 system 
and study its long-term stability and dynamics.

\section{Methodology}
\label{Dynamical fit}

Our orbital analysis for HD~59686 is based on the multi-dimensional $N$-body modeling scheme, which was previously applied 
to the 2:1 MMR exoplanet pairs around 
HD~82943 \citep{Tan2013}, HD~73526 \citep{Wittenmyer2014} and $\eta$~Ceti \citep{Trifonov2014}.
Briefly, we model the RV data using a Levenberg-Marquardt (L-M) $\chi^2$ 
minimization scheme, which performs an $N$-body fit by integrating the equations of motion
using the Gragg-Bulirsch-Stoer integration method \citep[see][]{Press}.
The output parameters from our fitting code
are the planetary and secondary star RV semi-amplitude ($K_{p,B}$), orbital period~($P_{p,B}$), eccentricity~($e_{p,B}$),
argument of periastron~($\omega_{p,B}$), mean anomaly~($M_{0 p,B}$), inclination~($i_{p,B}$) relative to the sky
plane, and ascending node ($\Omega_{p,B}$), as well as the RV offset (RV$_{\mathrm{off}}$).
All orbital parameters are the osculating ones in the Jacobi frame \citep[e.g.,][]{LeeM2003} at the first RV observational epoch, which is JD = 2451482.024.
Each fit comes with a reduced $\chi^2$ value ($\chi_{\nu}^2$), the residual $r.m.s.$ value,
and the 1$\sigma$ uncertainties of the adjusted parameters
obtained from the covariance matrix.

For HD~59686, we adopt a stellar mass of 1.92 $M_\odot$ and a stellar velocity jitter amplitude of 20~m\,s$^{-1}$.
\citet[][]{Ortiz2016} have shown that the Lick data of HD~59686 
are consistent with additional stellar radial velocity jitter of about 20~m\,s$^{-1}$.
The most likely reason for the notable RV noise in early K giants like HD~59686
are solar-like $p$-mode oscillations \citep{Barban, Zechmeister2008}, which 
have typical periods much shorter than the typical time 
sampling of our Lick data, and thus appear as scatter. 
Using the stellar parameters for HD~59686 from \citet[][]{Reffert2014} and the scaling relation from \citet{Kjeldsen2011},
we estimated a jitter amplitude of 16.1 $\pm$ 2.9~m\,s$^{-1}$, which agrees well with the observed jitter
of other K2~III giants in the Lick survey \citep[][]{Frink, Hekker, Trifonov2014, Reffert2014}.
Therefore, for our dynamical modeling we adopt a uniform a priori
stellar jitter value of 20~m\,s$^{-1}$, which we quadratically add\footnote{
Alternatively, the RV jitter could be fitted as a free parameter of the RV model \citep[e.g.][]{Baluev2009}.
The best double-Keplerian fit of HD~59686 optimized with an additional jitter term to the Lick data
yields a jitter value of $19.6_{-1.5}^{+1.8}$~m\,s$^{-1}$, which is 
consistent with the uniform jitter value of 20~m\,s$^{-1}$ adopted in this work.} 
to the total RV data error budget.

All dynamical fits in our study are further tested for long-term dynamical stability.
We integrate the orbits using a custom version of the Wisdom-Holman algorithm \citep[][]{Wisdom1991}, 
modified to handle the evolution of hierarchical systems consisting of massive bodies \citep{LeeM2003}.
The bodies are assumed to be point masses, and mutual collisions between them are not considered in defining system stability. 
We also neglect General Relativity and companion-star tidal effects during the simulations.
We integrate the individual fits for a maximum of 10~Myr, by adopting an integration time step equal to 1 day.
Our integration setup corresponds to more than $3 \times 10^5$ full binary orbits 
with about 300 steps per complete planetary orbit. 
We find that this setup is sufficient to resolve the planet's orbit with high resolution
and study the system's long-term stability.

We define the HD~59686 system as stable if during the integration the companion bodies remain in orbits
which do not deviate significantly from their initial best-fit configuration.
The system's stability depends primarily on the survival of the S-type planet. 
In most cases when the planet inclination is $30^\circ < i_p < 90^\circ$, the planet has relatively 
low mass to perturb the binary orbit significantly, and can well 
be approximated as a test particle in a two-body system.
However, we also test fits with $i_p < 30^\circ$, where the mass of the S-type body becomes quite large
as $\sin i_p$ gets smaller, and thus can significantly influence the binary orbit during the orbital evolution.
A simulation is terminated and the system is considered as unstable if at some point of the integration 
the semi-major axis $a_p$ or $a_B$ changes by more than $\pm$ 60\% from their initial values, or if $e_{p,B} > 0.95$.
 Particularly, when $e_p > 0.95$ and $a_p <  1.2$ AU, the planet periastron distance to the central star,
$q_p$, is well within the physical radius of the K giant ($q_p < R \approx$ 0.06 AU),
and the planet would collide with the star. 
Although these criteria for instability are somewhat arbitrary, our simulations show
that even small chaotic deviations in $a_p$ and $e_p$ quickly accumulate, 
and there are no cases where the orbits change significantly without exceeding these criteria.

\begin{table}[ht]
\begin{adjustwidth}{-4.0cm}{} 
\resizebox{0.69\textheight}{!}
{\begin{minipage}{1.1\textwidth}

\centering   
\caption{HD~59686 System Best Dynamical Fits}   
\label{table:orb_par_stable}      

\begin{tabular}{ lccccc}     

\hline\hline  \noalign{\vskip 0.7mm}      

\multicolumn{3}{c}{Coplanar edge-on prograde}           \\

\hline \noalign{\vskip 0.7mm}  

Parameter & HD~59686 Ab & HD~59686 B  \\

\hline\noalign{\vskip 0.5mm}

$K$  [m\,s$^{-1}$]                        & 137.0$_{-4.5}^{+3.6}$ ($\pm$3.4)      &  4012.6$_{-8.2}^{+9.9}$ ($\pm$20.6)        \\
$P$ [days]   			          & 299.1$_{-0.3}^{+0.3}$ ($\pm$0.3)     &  11696.4$_{-170.7}^{+209.2}$ ($\pm$196.4)       \\  
$e$                                       & ~~~~0.05$_{-0.02}^{+0.03}$ ($\pm$0.02) &  ~~~~0.730$_{-0.003}^{+0.004}$ ($\pm$0.003)   \\
$\omega$ [deg]                            & ~~121.1$_{-25.7}^{+28.5}$ ($\pm$28.7)  &  149.4$_{-0.1}^{+0.2}$ ($\pm$0.1)         \\   
$M_0$ [deg]                               & ~~299.5$_{-32.7}^{+20.9}$ ($\pm$28.4)  &  259.2$_{-1.5}^{+1.7}$ ($\pm$1.7)         \\  \noalign{\vskip 0.9mm} 
RV$_{ \mathrm{off}}$~[m\,s$^{-1}$]        & ~~248.6$_{-10.4}^{+13.5}$ ($\pm$12.5)  &                            \\ \noalign{\vskip 0.9mm} 
$i$ [deg]                                 & 90.0\tablenotemark{a}       & 90.0\tablenotemark{a}              \\ 
$\Omega$ [deg]                            & 0.0\tablenotemark{a}        & 0.0\tablenotemark{a} \\
$\Delta i$ [deg]                          & 0.0                 &                            \\ \noalign{\vskip 0.9mm} 
$a$ [AU]                                  & 1.089$_{-0.001}^{+0.001}$                &  13.611$_{-0.132}^{+0.163}$                     \\  
$m$ [$M_{\mathrm{Jup}}$]           & 6.97$_{-0.23}^{+0.18}$                &  558.41$_{-0.99}^{+1.20}$                    \\ \noalign{\vskip 0.9mm}
$r.m.s. $ [m\,s$^{-1}$]                   & 19.59               &                            \\ 
$\chi^2$                                  & 76.61               &                            \\  
$\chi_{\nu}^2$                            & 0.995               &                            \\
\noalign{\vskip 0.5mm}

\hline\hline\noalign{\vskip 1.2mm} 

\multicolumn{3}{c}{Coplanar edge-on retrograde}            \\

\hline \noalign{\vskip 0.7mm}  

Parameter & HD~59686 Ab & HD~59686 B  \\
\hline\noalign{\vskip 0.5mm}

$K$  [m\,s$^{-1}$]                        & 136.7$_{-4.4}^{+3.7}$ ($\pm$3.3)      &  4013.7$_{-7.7}^{+9.8}$ ($\pm$20.5)        \\
$P$ [days]   		        	  & 299.0$_{-0.3}^{+0.3}$ ($\pm$0.3)      &  11669.3$_{-147.0}^{+218.1}$ ($\pm$194.7)       \\  
$e$                                       & ~~~~0.05$_{-0.02}^{+0.03}$ ($\pm$0.02) &  ~~~~0.729$_{-0.003}^{+0.004}$ ($\pm$0.003)   \\
$\omega$ [deg]                            & ~~126.8$_{-24.5}^{+27.1}$ ($\pm$28.3)   &  149.4$_{-0.2}^{+0.2}$ ($\pm$0.1)         \\   
$M_0$ [deg]                               & ~~293.5$_{-51.5}^{+29.5}$ ($\pm$28.4)   &  258.9$_{-1.1}^{+2.6}$ ($\pm$1.7)         \\  \noalign{\vskip 0.9mm} 
RV$_{ \mathrm{off}}$~[m\,s$^{-1}$]        & ~~247.1$_{-9.3}^{+13.6}$ ($\pm$12.4)   &                            \\ \noalign{\vskip 0.9mm} 
$i$ [deg]                                 & 90.0\tablenotemark{a}        & 90.0\tablenotemark{a}              \\ 
$\Omega$ [deg]                            & 180.0\tablenotemark{a}       & 0.0\tablenotemark{a}               \\  
$\Delta i$ [deg]                          & 180.0                &                            \\ \noalign{\vskip 0.9mm} 
$a$ [AU]                                  & 1.089$_{-0.001}^{+0.001}$                   &  13.591$_{-0.145}^{+0.169}$                     \\  
$m$ [$M_{\mathrm{Jup}}$]           & 6.96$_{-0.23}^{+0.18}$                 &  558.46$_{-0.91}^{+1.21}$                    \\ \noalign{\vskip 0.9mm}
$r.m.s. $ [m\,s$^{-1}$]                   & 19.46               &                             \\ 
$\chi^2$                                  & 75.63             &                               \\   
$\chi_{\nu}^2$                            & 0.982             &                               \\
\noalign{\vskip 0.5mm}

\hline\hline\noalign{\vskip 1.2mm}

\multicolumn{3}{c}{Mutually inclined}            \\


\hline \noalign{\vskip 0.7mm}  

Parameter & HD~59686 Ab & HD~59686 B  \\
\hline\noalign{\vskip 0.5mm}

$K$  [m\,s$^{-1}$]                        & ~~130.1$_{-3.0}^{+3.1}$ ($\pm$26.2)  &  4020.0$_{-5.5}^{+6.2}$ ($\pm$182.1)        \\
$P$ [days]   			          & 300.5$_{-0.6}^{+0.2}$ ($\pm$0.5)     &  11398.3$_{-95.2}^{+204.0}$ ($\pm$1244.3)       \\  
$e$                                       & ~~~~0.08$_{-0.02}^{+0.02}$ ($\pm$0.02) &  ~~0.725$_{-0.002}^{+0.003}$ ($\pm$0.003)    \\
$\omega$ [deg]                            & ~~145.2$_{-17.6}^{+19.2}$ ($\pm$19.7)  &  149.8$_{-0.2}^{+0.2}$ ($\pm$4.7)          \\   
$M_0$ [deg]                               & ~~280.4$_{-19.9}^{+26.3}$ ($\pm$20.0)  &  256.4$_{-0.7}^{+2.4}$ ($\pm$21.0)         \\  \noalign{\vskip 0.9mm} 
RV$_{ \mathrm{off}}$~[m\,s$^{-1}$]        & ~~239.5$_{-9.0}^{+13.0}$ ($\pm$12.7)  &                             \\ \noalign{\vskip 0.9mm} 
$i$ [deg]                                 & 178.8$_{-0.5}^{+0.3}$ ($\pm$0.3)    & 86.4$_{-2.4}^{+2.4}$ ($\pm$3.3)             \\ 
$\Omega$ [deg]                            & ~~316.5$_{-17.7}^{+10.4}$ ($\pm$10.3)    & 0.0\tablenotemark{a}                \\ \noalign{\vskip 0.9mm}  
$\Delta i$ [deg]                          & 92.73                 &                            \\ \noalign{\vskip 0.9mm} 
$a$ [AU]                                  & 1.15$_{-0.02}^{+0.02}$                &  14.06$_{-0.18}^{+0.26}$                     \\  
$m$  [$M_{\mathrm{Jup}}$]                 & 359.22$_{-120.44}^{+114.33}$            &  618.36$_{-15.82}^{+18.45}$                    \\ \noalign{\vskip 0.9mm}
$r.m.s. $ [m\,s$^{-1}$]                   & 16.62               &                             \\ 
$\chi^2$                                  & 55.50             &                               \\ 
$\chi_{\nu}^2$                            & 0.750             &                               \\
\noalign{\vskip 0.5mm}

\hline\hline\noalign{\vskip 1.2mm}

\end{tabular}

\end{minipage}}

\end{adjustwidth}

\tablenotetext{a}{\small Fixed parameters.}
\tablecomments{\small $_{-0.0}^{+0.0}$ bootstrap uncertainties, ($\pm$0.0) covariance matrix uncertainties.}

\end{table}

\section{Best fits}
\label{best_fits}

\subsection{Edge-on prograde and retrograde fits}

\label{Prograde and retrograde}

\begin{figure}[btp]

\includegraphics[width=8cm]{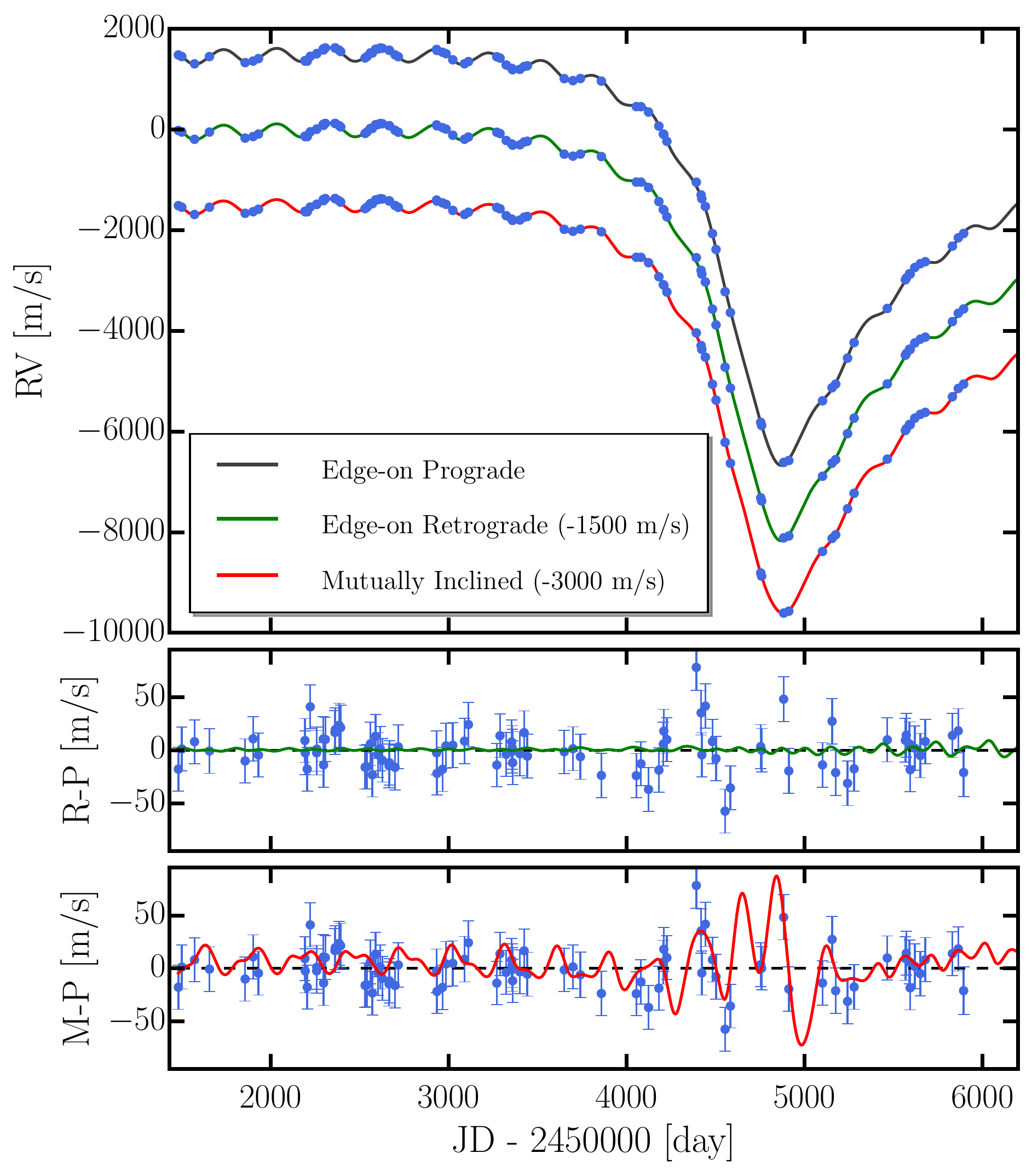}

\caption{Three best fit models to the Lick data ({\it blue points}).
In the upper panel,
the top curve ({\it black}) is the best edge-on coplanar prograde fit, while the middle and bottom curves are the best
edge-on coplanar retrograde ({\it green}) and mutually inclined ({\it red}) fits, offset vertically for
illustration purposes by $-$1500~m\,s$^{-1}$ and $-$3000~m\,s$^{-1}$, respectively.
Error bars include 20 m\,s$^{-1}$ added quadratically to the formal uncertainties to account for stellar jitter. 
The residuals of the best edge-on prograde fit are compared to
the difference between the prograde (P) fit and the best edge-on retrograde (R) or mutually inclined (M) fits in the lower two panels.
The difference between the prograde and retrograde fits is very small.
The mutually inclined fit models some data points with large residuals in the other orbital fits better, although
these data points lie in the relatively sparsely sampled epochs around the periastron passage of the binary orbit.
}
\label{diff_fit} 

\end{figure}

\begin{figure*}
\begin{center}$
\begin{array}{ccc} 

\includegraphics[width=6cm, height = 7cm]{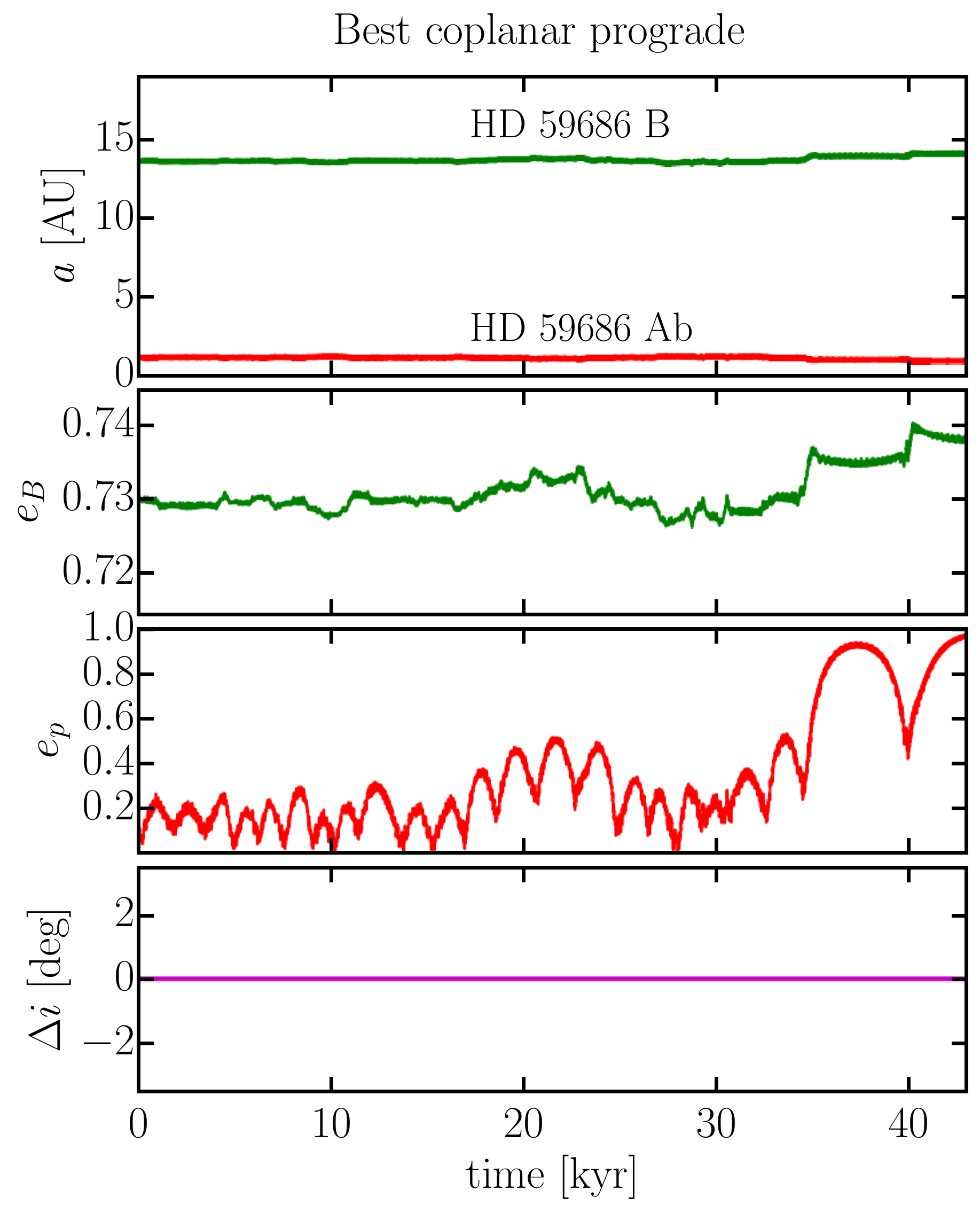} 
\includegraphics[width=6cm, height = 7cm]{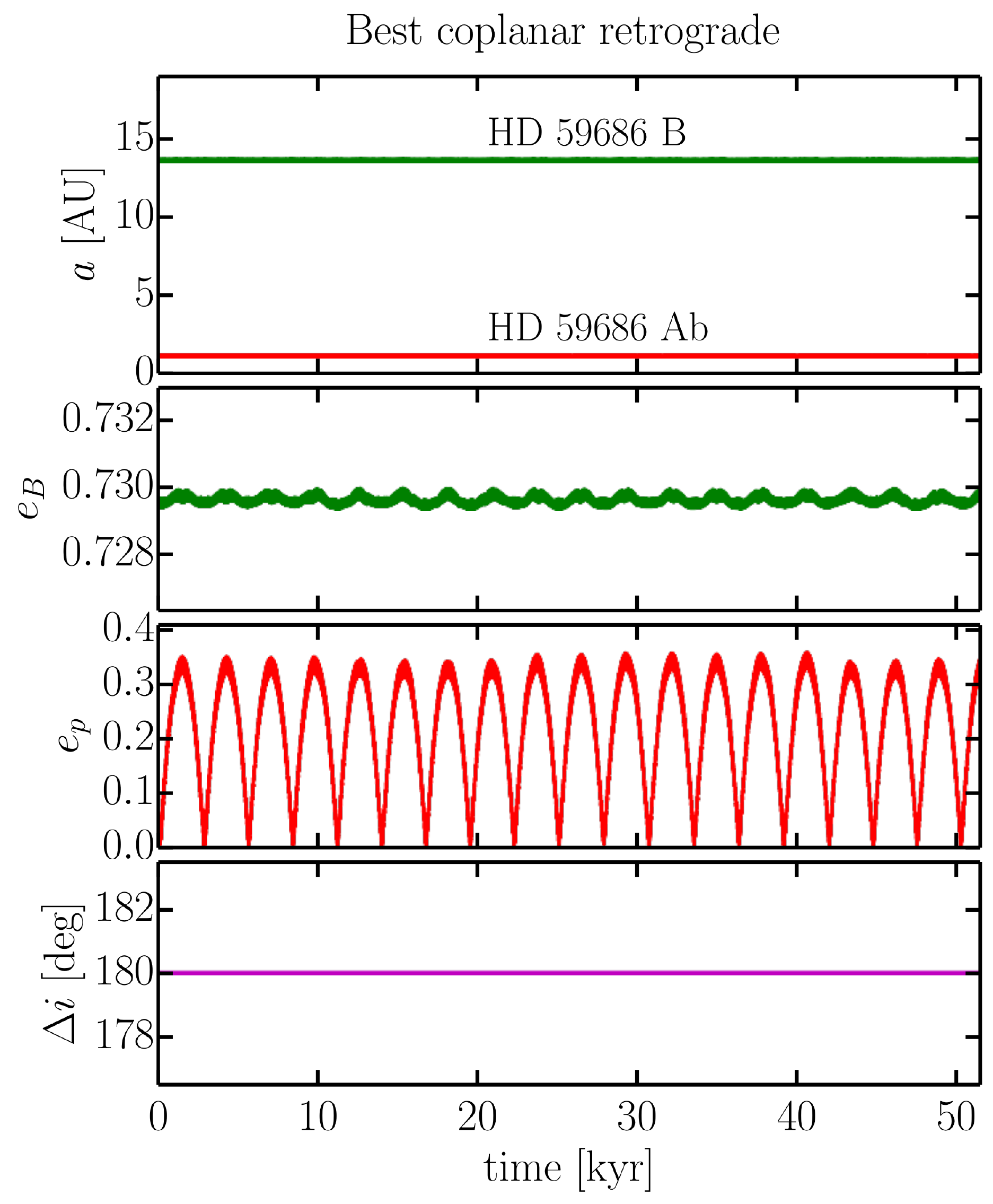} 
\includegraphics[width=6cm, height = 7cm]{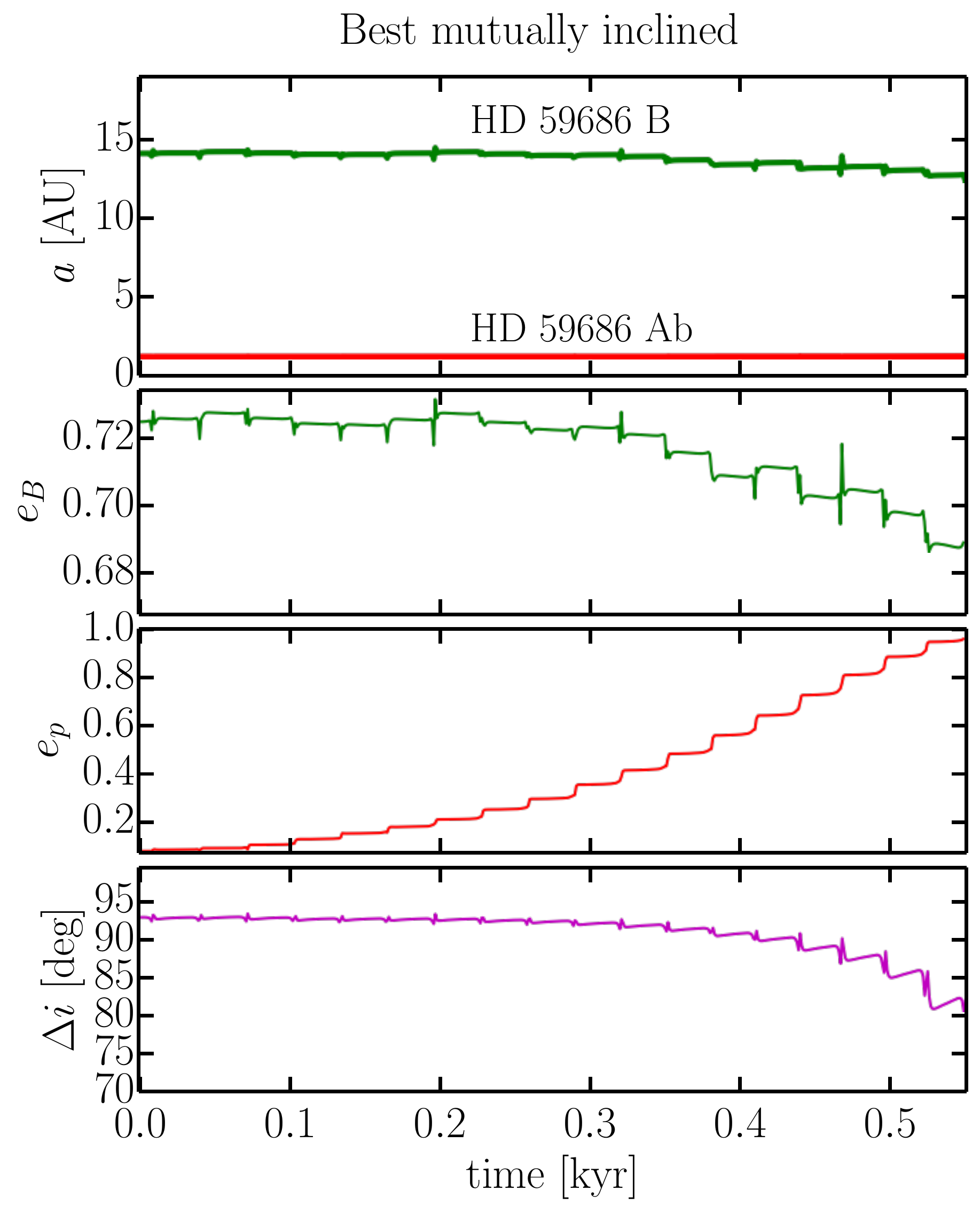}

\end{array} $

\end{center}

\caption{Semi-major axes, eccentricities, and mutual inclination evolution for the best coplanar, edge-on prograde and retrograde fits and 
the best mutually inclined fit. 
The best coplanar prograde fit ({\it left}) is unstable on short time scales of about 43 kyr,
when the planet eccentricity is excited to $e_p > 0.95$, leading to collision with the star.
The retrograde fit ({\it middle}) is stable during the 10 Myr test ($\sim$ 50 kyr shown).
In this fit, $e_p$ oscillates with large amplitude between 0 and 0.35, but the orbits remain well separated and stable.
The best mutually inclined configuration ({\it right}) does not survive even 600 years. 
The planet is initially on a nearly circular orbit, but with $\Delta i \approx$ 93$^\circ$ with respect to the binary plane. 
Due to the Lidov-Kozai effect, the planet eccentricity is quickly excited to $e_p > 0.95$.}
\label{best_fits_evol} 

\end{figure*}

The best coplanar and edge-on dynamical fit is generally consistent with the Keplerian fit shown in \citet{Ortiz2016}.
Our dynamical fit is close to a double Keplerian, since any significant gravitational perturbations on the planetary orbit 
(and thus on the induced RVs) are expected to be detected only after a few binary cycles, while the RV data currently 
cover only $\sim$ 40\% of one full binary orbit.
We first keep the orbital inclinations fixed at $i_p$ = $i_B$ = 90$^\circ$ and the difference 
between the lines of node $\Delta\Omega$ = $\Omega_p$ -- $\Omega_B$ = 0$^\circ$, which defines a planar and prograde configuration. 
The best fit in this orbital configuration has $\chi_{\nu}^2$~=~0.995 and leads to orbital elements of 
$P_p$~=~299.1~$\pm$~0.30 days, $e_p$~=~0.05~$\pm$~0.02, $a_p$ = 1.09 AU, and mass of $m_p$~=~6.97~$M_{\mathrm{Jup}}$ for the planet, 
and orbital elements of $P_B$~=~11696.4~$\pm$~196.4 days, $e_B$~=~0.73~$\pm$~0.003,  $a_B$ = 13.61 AU,
and secondary star mass of $m_B$~=~558~$M_{\mathrm{Jup}}$ for the binary.
The full set of orbital elements and their bootstrap (see \S\ref{Statistical}) and covariance matrix
estimated uncertainties are given in Table \ref{table:orb_par_stable}, while the actual fit to the data (black curve in the upper panel)
and its residuals are illustrated in Figure~\ref{diff_fit}. 
The long-term evolution of the orbital semi-major axes and eccentricities are shown in Figure~\ref{best_fits_evol} (left panel).
According to our stability criteria, this fit is stable only for about 42 kyr, before the planet collides with the star.
A close examination of this fit indicates that the orbits show large variations in $e_p$
and small, but chaotic variations in $e_B$.
Eventually the secondary companion excites the planet eccentricity above $e_p > 0.95$, which interrupts our integration.

\begin{figure}[btp]

\begin{center}$
\begin{array}{ccc} 
\includegraphics[width=9cm]{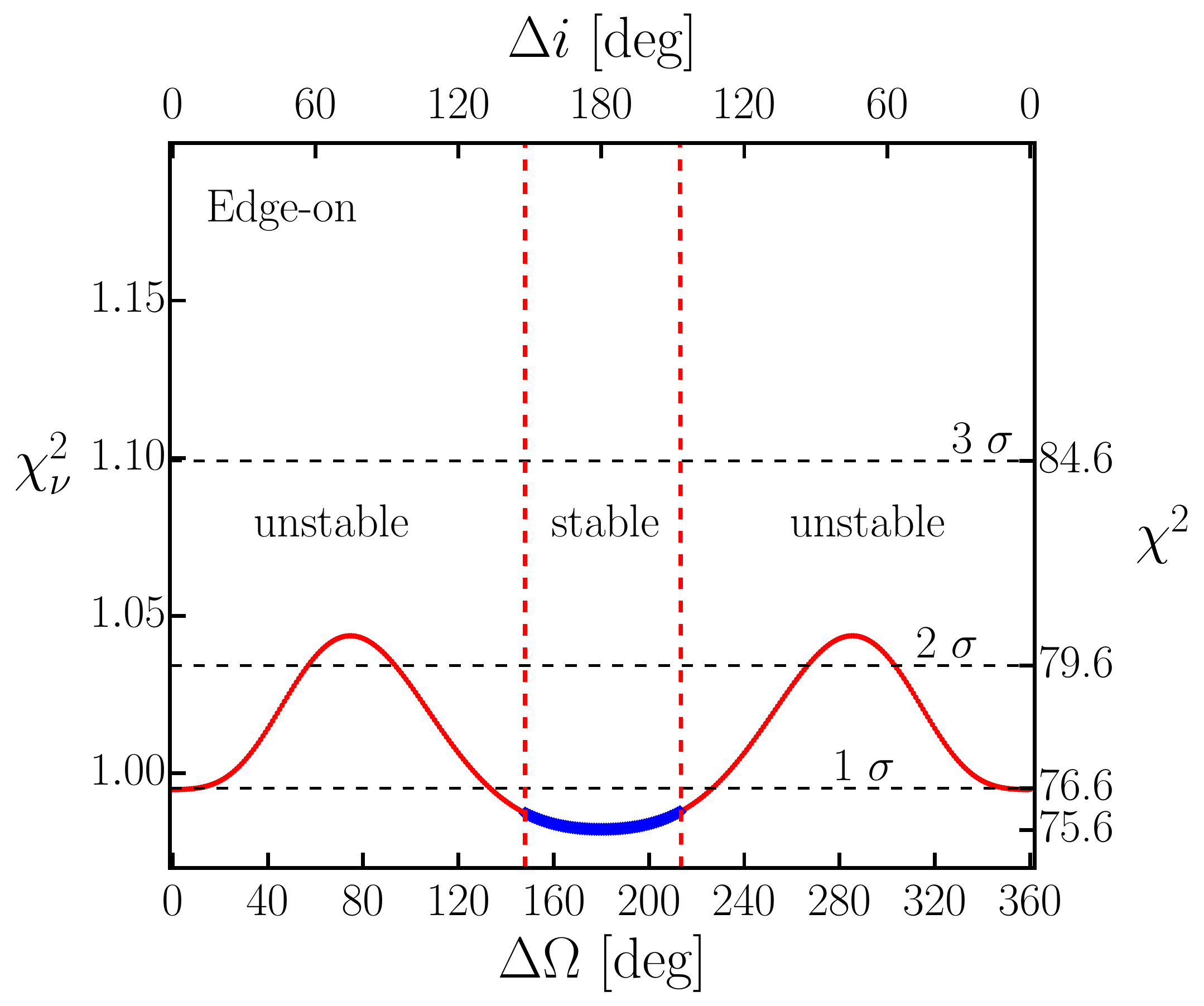}\\
\end{array} $

\end{center}

\caption{Edge-on ($i_p$ = $i_B$ = 90$^\circ$), but mutually inclined fits of HD~59686.
The mutual inclination angle $\Delta i$ in edge-on orbits comes from $\Delta\Omega$ = $\Omega_p$ - $\Omega_B$. 
The $\Delta\chi^2$ confidence levels in terms of 1$\sigma$, 2$\sigma$ and 3$\sigma$ are drawn for the       
$\chi^2$ minimum, which is at edge-on and retrograde orbits ($\Delta i$ = 180$^\circ$).
All fits between $\Delta i$ $\sim$ 145$^\circ$ and 180$^\circ$ ({\it blue}) are stable. These fits also have 
better quality when compared to the prograde, polar and near-polar fits, which are unstable. 
}
\label{incl_reduced_chi_both_3} 
\end{figure}

Since the best coplanar and prograde fit is unstable, we test how the fit quality and stability change if we allow non-coplanar orbits. 
We simplify this test by keeping the binary on an edge-on orbit with fixed $i_B$ = 90$^\circ$ and $\Omega_B$ = 0$^\circ$.
For the planet we also fix the inclination at $i_p$ = 90$^\circ$, but 
we systematically vary $\Omega_p$ between 0$^\circ$ and 359$^\circ$ with a step of 1$^\circ$.
Thus, in this test we keep the companion masses at their minimum, while 
the mutual inclination comes only from the difference between 
the longitudes of the ascending nodes $\Delta\Omega$ = $\Omega_p$ -- $\Omega_B$ following the expression:
\begin{equation}
\begin{split}
\Delta i = \arccos[\cos(i_p)\cos(i_{B})+\sin(i_p)\sin(i_{B})\cos(\Delta\Omega)] .
\end{split}
\label{eq:deltai}
\end{equation}
Figure~\ref{incl_reduced_chi_both_3} shows the results from this test. 
We plot the quality of the mutually inclined fit in terms of $\chi_{\nu}^2$ ($\chi^2$) as a function of $\Delta\Omega$ ($\Delta i$).
With horizontal dashed lines are shown the 1$\sigma$, 2$\sigma$ and 3$\sigma$ confidence levels according to $\Delta\chi^2$.
The best-fit in Figure~\ref{incl_reduced_chi_both_3} appears at $\Delta i$ = 180$^\circ$, which is again a coplanar, but retrograde planet orbit.
The best coplanar prograde fit has $\chi_{CP}^2$ = 76.61, while the best coplanar retrograde fit has $\chi_{CR}^2$ = 75.63,
resulting in $\Delta\chi^2$ = 0.98. This difference is slightly below the 1$\sigma$ limit, and thus 
 the retrograde fit does not represent a significant improvement to our model.
In Figure~\ref{incl_reduced_chi_both_3} most of the edge-on fits with $\Delta i$ $<$ 145$^\circ$ are above 1$\sigma$ from the best fit and are unstable (red dots),
while all fits with $\Delta i$ between 145$^\circ$ and 180$^\circ$ are within 1$\sigma$ and are stable (blue thick line) for at least~10~Myr.

Figure~\ref{best_fits_evol} (middle panel) shows a $\sim$ 50 kyr time span of the orbital evolution for the best coplanar retrograde fit.
The semi-major axes $a_p$ and $a_B$ are nearly constant during the stability test.
The planet eccentricity $e_p$ oscillates with a large amplitude between 0 and 0.35,
but the system remains stable, with the bodies well separated from each other. 
Interestingly, the mean period ratio of this stable retrograde fit is $P_B/P_p \approx$ 39, but the system is not in 39:1 MMR, as none of the resonance angles associated with
the 39:1 MMR are librating.
For the $n:1$ MMR, the resonance angles are

\begin{equation}
\theta_{m=1,n} =  \lambda_p  - n\lambda_B + (m-1)\varpi_p  - (m-n)\varpi_B, \\
\label{eq:theta}
\end{equation}

\noindent
where $n$ is positive for prograde motion and negative for retrograde motion,
$\varpi_{p,B}$ are the longitudes of periastron and $\lambda_{p,B}$ are the mean longitudes.
All fits with $\Delta i$ between 145$^\circ$ and 180$^\circ$ have similar behavior for $a_p$, $a_B$, $e_p$, and $e_B$,
while $\Delta i$ oscillates with small amplitude around the initial fitted value.
None of them seems to be locked in a MMR.

The most likely reason for the wider stable region for the retrograde orbits is that
the individual MMR are of higher order for retrograde than prograde orbits.
As demonstrated in \citet{Morais2012}, at
the same $n:q$ mean-motion ratio, the MMR is of order $n - q$ for prograde versus order $n + q$ for retrograde orbits,
 which for the latter results in much narrower MMR libration widths and thus smaller phase-space overlap 
of neighboring MMR where the planet would be likely unstable.
The findings of \citet{Morais2012}, however, were restricted to the dynamics of S-type planets in circular binary systems, 
while the dynamics of the planet in the highly eccentric HD 59686 binary is far more complex. 
A more detailed analysis of resonance width and overlap for 
prograde and retrograde orbits using the formalism developed by \citet{Mardling2008}
in the context of the HD 59686 system will be presented in a future paper (Wong \& Lee, in preparation).


\subsection{Inclined coplanar fits -- constraining $\sin i$}
\label{Inclined coplanar}

Both prograde and retrograde edge-on best-fits suggest a coplanar configuration.
Therefore, as a next step we test how the fit quality and stability 
for both configurations depends on the inclination $i$ (measured from the plane of the sky).
For the prograde geometry we fix $\Delta\Omega$ = 0$^\circ$ and $i_p$ = $i_B$. 
We systematically vary $i_p$ and $i_B$ from 90$^\circ$ to 5$^\circ$ with a decreasing step of 1$^\circ$,
and thus we gradually increase the companion masses by a factor of approximately $\sin i$.
The same test is done  for the retrograde fits, which
are constructed by keeping $\Delta\Omega$ = 180$^\circ$, $i_B$ = 180$^\circ$ $-$ $i_p$.
and varying $i_p$ from 90$^\circ$ to 5$^\circ$ with a step of 1$^\circ$.
According to Equation (\ref{eq:deltai}), the prograde fits have a mutual inclination of
$\Delta i$ = 0$^\circ$ and the retrograde fits $\Delta i$ = 180$^\circ$.

\begin{figure}[btp]

\begin{center}$
\begin{array}{ccc} 
\includegraphics[width=9cm]{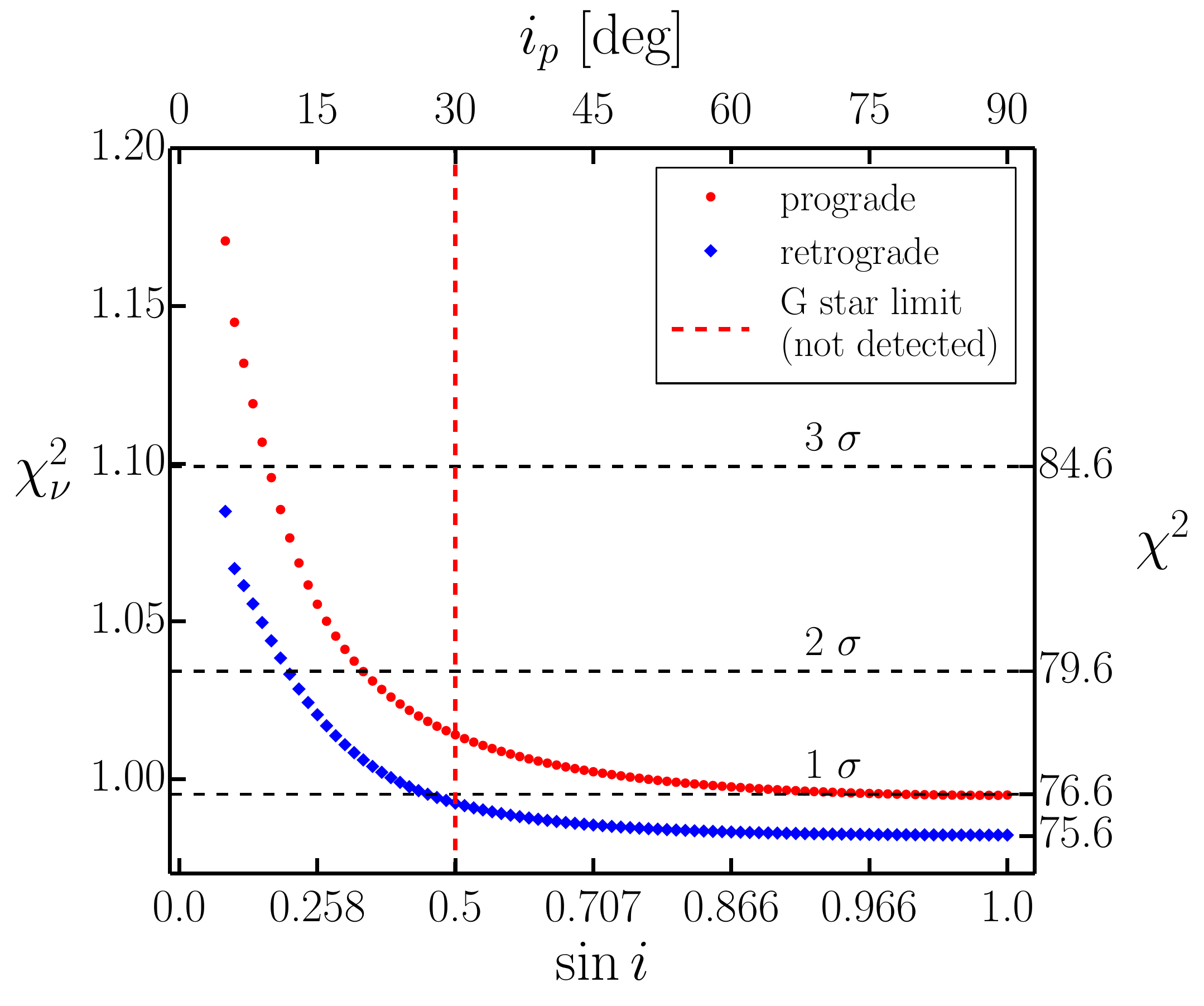} \\
\end{array} $ 

\end{center}

\caption{Comparison of the quality of coplanar prograde and retrograde fits as a function of the inclination $i$. 
Retrograde configurations have better $\chi_{\nu}^2$ than prograde ones and 
are all stable for 10 Myr, while none of the prograde fits is long-term stable.
In both cases, $\chi_{\nu}^2$ is minimum at $i$ = 90$^\circ$ (edge-on) and increases only slowly with decreasing $i$. 
The $\Delta\chi^2$ confidence levels (1$\sigma$, 2$\sigma$ and 3$\sigma$)
are obtained from the best retrograde fit as in Figure~\ref{incl_reduced_chi_both_3}.
The red dashed line marks $i = 30^\circ$, below which the 
secondary must be at least a G dwarf mass star, which would have been detected with 
LBT angular differential imaging \citep{Ortiz2016}.
} 
\label{incl_reduced_chi_retro} 
\vspace*{1cm}
\end{figure}

The results from this systematic test are illustrated in Figure~\ref{incl_reduced_chi_retro}, which
shows a comparison between prograde and retrograde dynamical fits as a function of $\sin i$. 
The $\chi_{\nu}^2$ minimum for both prograde and retrograde cases is at $\sin i$~=~1, which corresponds to the same 
best fits presented in Table~\ref{table:orb_par_stable} and Figure~\ref{incl_reduced_chi_both_3}.
The $\Delta\chi^2$ confidence levels (1$\sigma$, 2$\sigma$ and 3$\sigma$) in Figure~\ref{incl_reduced_chi_retro}
are measured from the best retrograde fit, and represent the same confidence levels as shown in Figure~\ref{incl_reduced_chi_both_3}.
Clearly, our $N$-body fits can only weakly constrain the orbital inclination from the RV data.
Overall, the retrograde fits have better $\chi_{\nu}^2$ values, but in both configurations
the fits are gradually becoming worse for lower $\sin i$, and thus higher planet and secondary star masses. 
For retrograde fits down to 2$\sigma$ the orbital inclination can be 
between 10$^\circ$ and 90$^\circ$, while for prograde fits the inclination can be between $20 ^\circ$ and 90$^\circ$.
In both configurations, however, the inclination is unlikely less than $30^\circ$,
as the secondary star would then be at least a G-type main-sequence star with about twice the minimum mass.
Such a stellar companion should have been detected by \citet{Ortiz2016} 
via LBT angular differential imaging,
but since it was not, we assume $\sin i_B$ = 0.5 as a lower limit.

When it comes to stability, all retrograde coplanar fits in Figure~\ref{incl_reduced_chi_retro}
are stable for 10 Myr, including those at very low inclinations, while none of the prograde fits is long-term stable.

\subsection{Mutually inclined fits -- the global minimum}
\label{Mutually inclined best fit}

Finally, in our dynamical modeling of HD~59686's RV data, we allow
non-edge-on mutually inclined orbits by fitting independently 
$i_p$ and $i_B$ in the range between 0$^\circ$ and 180$^\circ$ and $\Delta\Omega$ between 0$^\circ$ and 360$^\circ$. 
In this way, we allow our fits to adopt a large range of companion masses
and we cover all possible orbital alignments. 

Our  mutually inclined best-fit has a strong minimum with $\chi_{\nu}^2$~=~0.75, yielding
for the inner companion $P_p$~=~300.5~$\pm$~0.5 days, $e_p$~=~0.08~$\pm$~0.02,
$a_p$ = 1.15 AU, and for the outer companion $P_B$~=~11398.3~$\pm$~1244.3 days, $e_B$~=~0.725~$\pm$~0.003, 
$a_B$ = 14.06 AU (see Table \ref{table:orb_par_stable}).
Remarkably, this fit suggests that the inner companion has nearly face-on orbit with well constrained $i_p$ = 178.8 $\pm$ 0.3$^\circ$. 
This means that the inner companion is no longer a planet 
but a stellar mass companion with $m_p$ = 359$M_{\mathrm{Jup}}$ ($= 0.34 M_\odot$)
forming an inner binary pair with the K~giant.
The outer companion has nearly edge-on orbit with $i_B$ = 86.4 $\pm$ 3.3$^\circ$
and mass $m_B$ = 618~$M_{\mathrm{Jup}}$ ($= 0.59 M_\odot$). 
The difference between the ascending nodes is $\Delta \Omega$ = 316.5 $\pm$ 10.3$^\circ$, 
and according to Equation (\ref{eq:deltai}) this leads to $\Delta i$ = 92.7 $\pm$ 3.3$^\circ$.
We achieve practically the same fit (within errors) with $\chi_{\nu}^2$~=~0.75 
at $i_p$ = 1.15 $\pm$ 0.6$^\circ$, $i_B$ = 93.4 $\pm$ 2.5$^\circ$ and $\Delta \Omega$ = 43.7 $\pm$ 12.3$^\circ$, which is a 
mirror image of the above orbital configuration.

The $\chi^2$ value for this mutually inclined fit is 55.5, which is much 
lower than the best coplanar retrograde fit with $\chi_{CR}^2$ = 75.6.
This fit, however, has three additional fitting parameters compared 
to the edge-on coplanar fits, which must be taken into account when testing for significance.
Following the $\Delta\chi^2$ approach, we assume that $i_p$, $i_B$ and $\Delta\Omega$ 
are systematically adjusted, while the rest of the orbital parameters are fitted by our $N$-body model.
The $\Delta\chi^2$ confidence intervals in this case obey the $\chi^2$ distribution for 3 degrees of freedom.
The difference between the fits is $\Delta\chi^2$ = $\chi_{CR}^2$ $-$ $\chi_{MI}^2$ = 20.1, 
suggesting that the best coplanar retrograde fit is between 3$\sigma$ and 4$\sigma$ 
worse than the mutually inclined fit, and thus the latter represents a significant model improvement.

Since the coplanar model with $p_1$ = 11 fitting parameters is ``nested'' within the mutually inclined model 
with $p_2$ = 14 parameters, another way to test the significance is the use of the 
$F$-test and determine the $F$-value following \citep{Bevington2003}:
\begin{equation}
F = \frac{(\chi_{CR}^2 - \chi^2) / \zeta_1  }{\chi^2 / \zeta_2} = \frac{\Delta\chi^2 / \zeta_1  }{\chi_{\nu}^2} = 8.95 ,
\end{equation}
where $\zeta_1$ = $p_2-p_1$ is the number of additional parameters being tested, 
$\zeta_2$ = $n-p_2$ is the number of degrees of freedom for the best mutually inclined model,
with $n$ the number of data points. 
For $F = 8.95$, the probability for model improvement is $p$ = 0.000039, which is much lower than our adopted 
cut-off value of $\alpha$~=~0.01, meaning that the null hypothesis is successfully rejected.
Thus we conclude that the mutually inclined fit is indeed better when compared to the coplanar edge-on model.

This significant model improvement is intriguing and deserves a closer look.
First, it should be noted that the $\Delta\chi^2$ and $F$ tests only
work well for Gaussian errors and models that are linear in the
parameters (or could be linearized in the uncertainty region of the
parameters due to large enough sample size) \citep{Press}, which we do
not have when we apply an $N$-body dynamical fit to the existing RV
data for HD~59686.

Dynamical fitting of RV data consistent with two or more companions can be in principle sensitive to the true companion masses, but 
this has been proven to be very challenging even for the most extensively 
studied multi-planet systems \citep[see][]{Bean2, Correia2010,Nelson2016}.
The critical requirements to measure successfully mutual inclinations are:
(1) high RV precision, (2)  low velocity jitter, typically on the order of at most a few m\,s$^{-1}$, 
(3) large set of RV data covering many orbital cycles, and (4) 
the signal discrepancy between the minimum mass coplanar fit and 
the mutually inclined fit must be larger than the RV noise. 
In this context, we note that the available Lick data for HD~59686 do not satisfy these criteria,
with only 88 RVs (with precision of 5--9~m\,s$^{-1}$) distributed over 11 years covering only $\sim$ 40 \% of the outer binary orbit,
and RV jitter of $\sim$ 20 m\,s$^{-1}$. 
Thus it is unlikely that we would be able to tightly constrain the true companion masses (through $\Delta i, i_p, i_B$).
 
In the bottom panel of Figure~\ref{diff_fit}, the RV residuals to the coplanar prograde fit are shown, and 
over-plotted (red curve) is the difference of the mutually inclined
model from the coplanar prograde one. 
Clearly, some of the outliers present in the coplanar prograde case are well modeled 
by the mutually inclined model with three additional fitting parameters.
These outliers, however, lie in the relatively
sparsely sampled epochs between JD = 2454200 and 2455200, which unfortunately coincides
with the outer binary periastron passage, when the RV signal changes rapidly.
Perhaps, the signal would be validated if we had more RV data following the mutually inclined fit prediction 
at that orbital phase, but currently it is fair to conclude that we could be fitting noise rather than a true signal.

\begin{figure*}[btp]
\begin{center}$
\begin{array}{cc} 

\includegraphics[width=18cm]{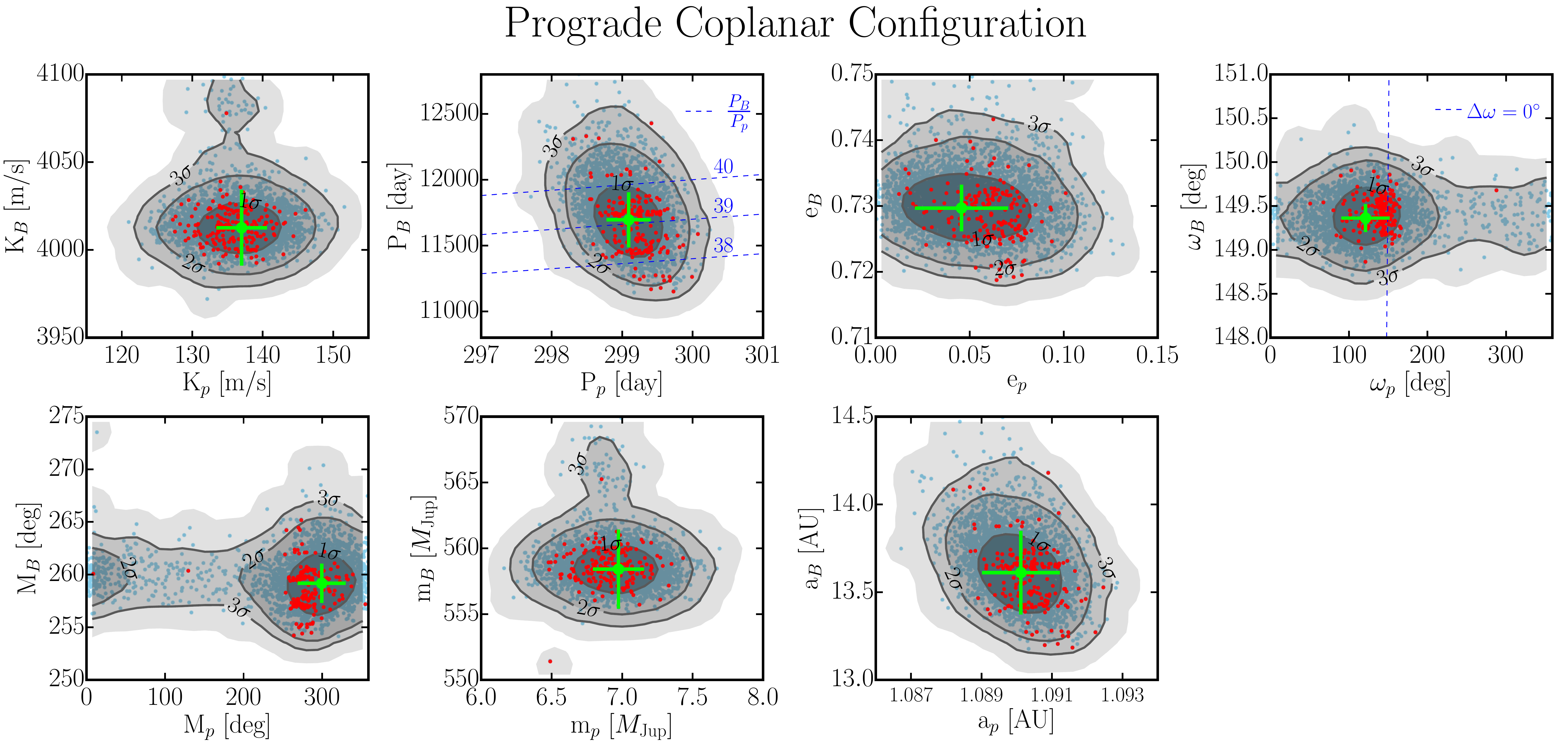}\\
\includegraphics[width=18cm]{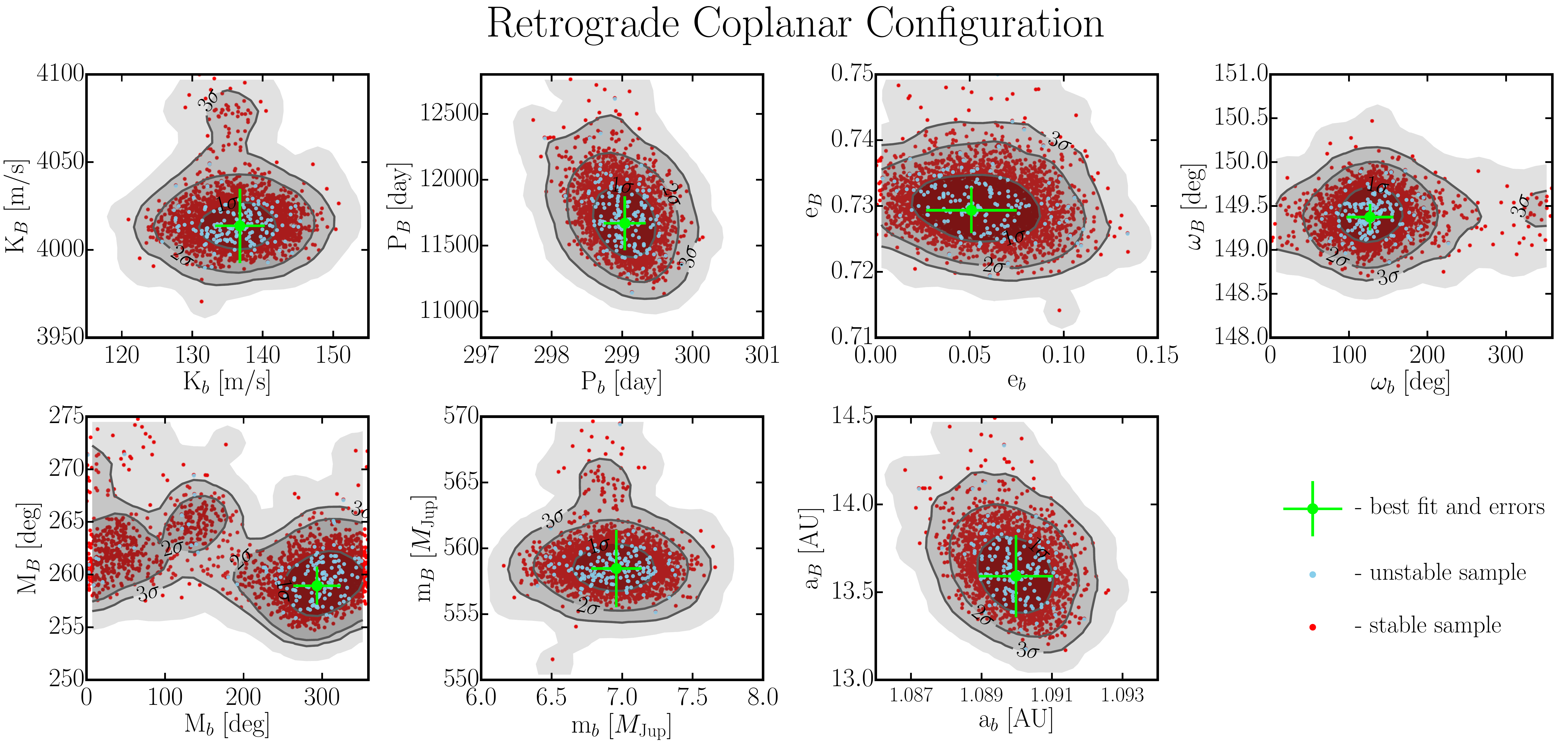}\\	
\end{array} $
\end{center}

\caption{Distribution of orbital parameters from dynamical fits to the bootstrapped RV data sets for the
 edge-on prograde (upper panels) and retrograde (lower panels) configurations.
Best-fit values and covariance matrix errors based on the original data set
are consistent  with the 1$\sigma$ confidence level of the bootstrap analysis. 
Only $\sim$4\% of the prograde fits within the 68.3\% confidence region are stable, while
97\% of the retrograde fits are stable for at least 10 Myr.  
The prograde stable fits are located at initial
$\omega_p$ $\approx$ 145$^\circ$, $e_p$ $\approx$ 0.06$-$0.08 and $P_B/P_p \neq 39$ or 40.
}   

\label{bootstrap} 
\end{figure*}

\begin{figure*}[btp]
\begin{center}$
\begin{array}{cc} 

\includegraphics[width=18cm]{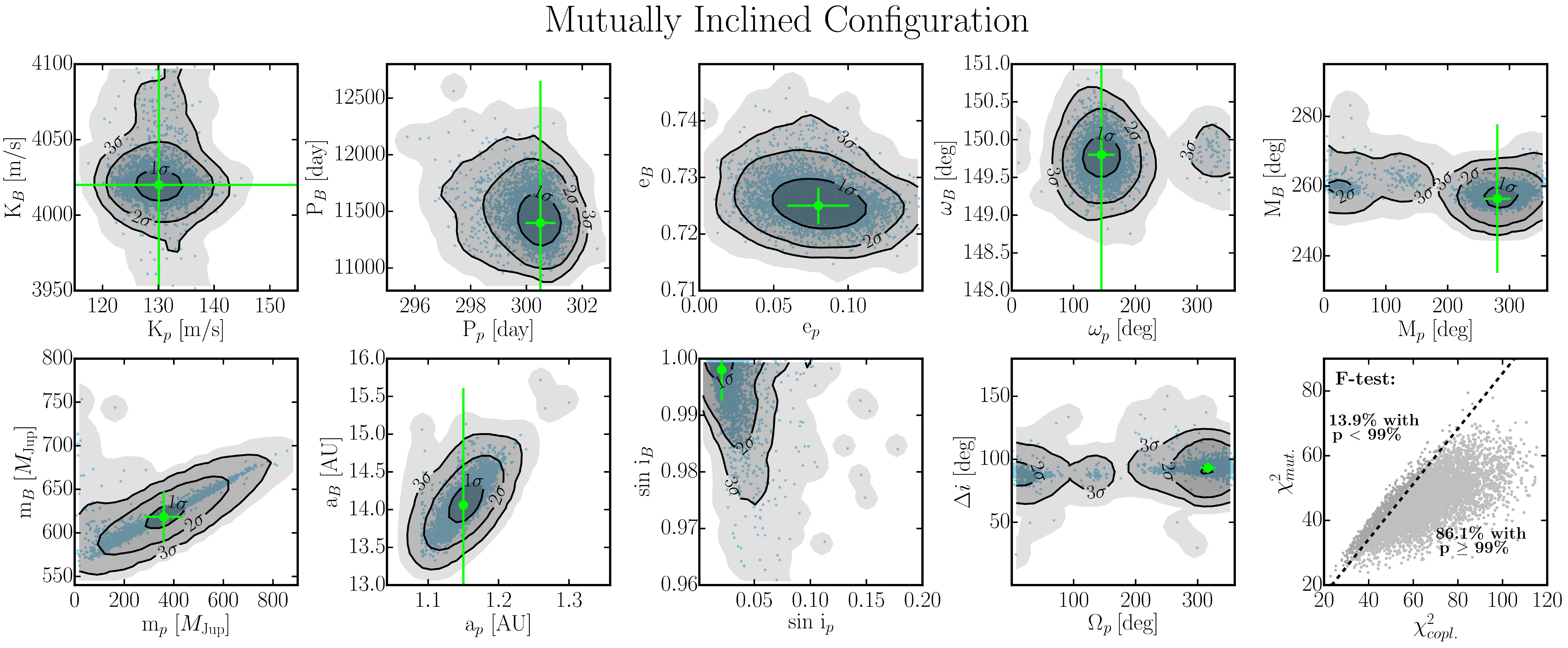}\\	
\end{array} $
\end{center}

\caption{Same as Figure~\ref{bootstrap}, but for mutually inclined configurations.
Best-fit values and errors from the covariance matrix are consistent with the bootstrap distributions at the 68.3\% 
(1$\sigma$) confidence level, except for larger covariance matrix errors for some binary parameters. 
$F$-test shows that 86.1\% of the mutually inclined fits present a significant improvement over the 
prograde fits, but none of them is stable. 
}   

\label{bootstrap2} 
\end{figure*}

In addition, this mutually inclined best fit is extremely unstable.
The inner companion has a nearly circular and highly inclined (polar-like) orbit with respect to the outer binary plane. 
Such orbits are potentially unstable due to the Lidov-Kozai effect \citep{Lidov1962, Kozai1962},
which leads to periodic exchanges between $\Delta i$ and $e_p$.
Following the analytic expression given in \citet{Takeda2005},
we can estimate the inner companion maximum eccentricity 
$e_{p, max}$ that can be reached through the Lidov-Kozai cycle:
\begin{equation}
e_{p, max} = \sqrt{1 - (5/3)\cos^2(\Delta i)}
\end{equation}
With $\Delta i$ = 92.7$^\circ$ obtained from the best fit, $e_{p, max}$ $\approx$ 0.998, which exceeds our stability criterion. 
As expected, our direct numerical integration (right panel of
Figure~\ref{best_fits_evol}) shows that the best mutually inclined fit
does not survive even 600 years.
The inner companion is quickly excited to $e_p > 0.95$ and it eventually collides with the K~giant.
Obviously, such a highly inclined stellar triple is very unstable,
which seems to be a good argument against this solution.

 \section{Bootstrap statistics}
\label{Statistical}

Our goal in this section is to obtain parameter estimates and confidence regions 
based on the  empirical distribution of constructed orbital parameters using bootstrap re-sampling.
We analyze the distribution of the adjustable fitting parameters around the best-fit 
by randomly drawing RV data points with replacement \citep[e.g.,][]{Efron1979, Press, Tan2013}
and perform  a dynamical fit to each RV data set obtained in this way.

We create a total of $n$ = 5000 bootstrapped data sets and we fit 
each sample with strictly coplanar edge-on (prograde and retrograde) and mutually inclined configurations.
All fits to the  bootstrapped data sets are integrated for a maximum of 10 Myr and their stability is examined.
The 1$\sigma$ confidence levels from the  distributions are used 
to estimate the asymmetrical best-fit parameter uncertainties. 
These estimates are listed in Table~\ref{table:orb_par_stable}, along with the best-fit covariance matrix errors.
For the mutually inclined fits, a bootstrapped sample is rejected if the fit 
suggests $i_B < 30^\circ$ or $i_B > 150^\circ$ (i.e., $\sin i_B < 0.5$).
This ensures that our samples are consistent with a reasonable secondary companion mass
in the range $m_B \approx 0.53$ to about 1.1$M_\odot$,
and thus are consistent with the observational constraints given in \citet{Ortiz2016}.
The total fraction of fits with highly inclined ($\sin i_B < 0.5$)
outer companion is $\sim$ 6\% of the total constructed bootstrapped RV data sets.
In addition, $\sim$~1.5~\% of the fits are unable to converge when we try to fit a mutually inclined 
configuration, meaning that in order to get  5000 fits we have to create
a total of $\sim$ 5400 bootstrapped RV data sets.

Figure~\ref{bootstrap} shows the  results from the bootstrap analysis for the edge-on ($i = 90^\circ$) 
prograde ($\Delta i = 0^\circ$) and retrograde ($\Delta i = 180^\circ$) configurations.
In each panel we illustrate the distribution of planet versus binary orbital elements 
($K$, $P$, $e$, $\omega$, $M_0$), semi-major axes $a$, and dynamical masses $m$.
Solid contours show the 1$\sigma$, 2$\sigma$ and 3$\sigma$ (68.27\%, 95.45\%, 99.73\%)
confidence levels from the two-dimensional parameter distributions.
In all panels, the green dot represents the position of the best-fit values from the 
prograde or retrograde dynamical fit to the original data set, while the green error bars are the estimated 
uncertainties from the covariance matrix (see Table~\ref{table:orb_par_stable}). 
With blue dots we mark the unstable configurations, while with red points we 
show the configurations which survive for 10 Myr.

 Clearly, the distributions of orbital elements for the prograde and retrograde configurations are very similar.
For both configurations, the covariance matrix errors estimated from the original data and the 
1$\sigma$ (68.3\%) confidence region from the bootstrap analysis are roughly consistent with each other. 
The main difference between the configurations comes from the stability results. 
We find that $\sim$ 97\% of the retrograde configurations are stable for 10 Myr,
while for the prograde case this number is only $\sim$ 4\%. 
We note, however, that the stable prograde configurations fall mostly 
within the 1$\sigma$ confidence region and from their distribution 
we can identify potentially stable regions of the parameter space.
As can be seen in Figure~\ref{bootstrap}, the stable fits seem to cluster 
around $\omega_p \approx \omega_B \approx$ 145$^\circ$, $M_{0,B}$ $\approx$ 257$^\circ$ and 259$^\circ$,
$e_p$ $\approx$ 0.05 to 0.07, and a few discrete values in $P_B$ which avoid initial integer period ratios of 39 and 40.
This result implies that a small, but robust set of stable prograde fits does exist. 
Therefore, we cannot eliminate the possibility that the HD~59686 system is in a stable prograde configuration.

The results of the bootstrap analysis for the mutually inclined configuration are shown in Figure~\ref{bootstrap2}.
In particular, we aim to quantitatively estimate the inclination distribution ($i_p, i_B, \Delta i$) 
and see how often polar-like S-type companion orbits will occur in the resampled data.
Therefore, in Figure~\ref{bootstrap2} we introduce three additional panels: 
$\sin i_p$ - $\sin i_B$, $\Delta i$ - $\Delta\Omega$ and a comparison between 
the mutually inclined $\chi_{mut}^2$ and prograde coplanar $\chi_{copl}^2$ values.
The last of these panels also shows the $F$-test probability for significant  improvement  when  three additional parameters are added.
We find that $\sim$ 86.1\% of the fits lead to significant improvement, applying our chosen threshold of $\alpha$ = 0.01.

The orbital parameter distributions are wider than in the coplanar cases
but seem to agree with the best-fit errors.
The binary orbit covariance matrix errors are sometimes larger than 
the bootstrap 1$\sigma$ contours, but in general the distribution peak is consistent with the best-fit estimate.
The inner companion inclination $i_p$ is found mostly at lower values, 
driving $m_p$ towards brown dwarf and star like masses.
On the other hand, within 3$\sigma$, $i_B$ stays between 75$^\circ$ and 105$^\circ$ (i.e., $\sin i_B$ between 1 and 0.96),
leading to $m_B$ similar to those of the edge-on cases.
The distribution of $\Delta i$ clusters around the best-fit
and is well constrained around $\sim$ 90$^\circ$, leading to nearly perpendicular triple-star orbits. 
We suspect, however, that the distribution in $\Delta i$ might be the result of a model degeneracy. 
By randomly scrambling the data with repetition, the possibly problematic outliers with sparse cadence are not always removed (see Figure~\ref{diff_fit} and
discussion in Section \S\ref{Mutually inclined best fit}).
Even worse, for some bootstrapped data sets, we repeat these points, while removing data points with lower residuals.
We find that none of the mutually inclined fits based on bootstrap analysis and shown in  Figure~\ref{bootstrap2} is long-term stable.
The average bootstrap survival time is only a few hundred years before the inner binary pair collides,
which is consistent with the orbital evolution of the best mutually inclined fit to the original data.

\section{Parameter grid search }
\label{systematic}

A detailed picture of the dynamical properties of the HD~59686 system
can also be assessed using the parameter grid-search technique \citep[e.g.,][]{Lee2006}.
We systematically vary a pair of orbital parameters, which we then keep fixed
while the rest of the parameters in the model are adjusted to minimize $\chi^2$.
This method allows us to systematically inspect the multi-dimensional parameter space around the best fit
and study the properties and long-term stability of nearby fits.

\begin{figure}[btp]
 \begin{center}$
\begin{array}{ccc} 

\includegraphics[width=9cm]{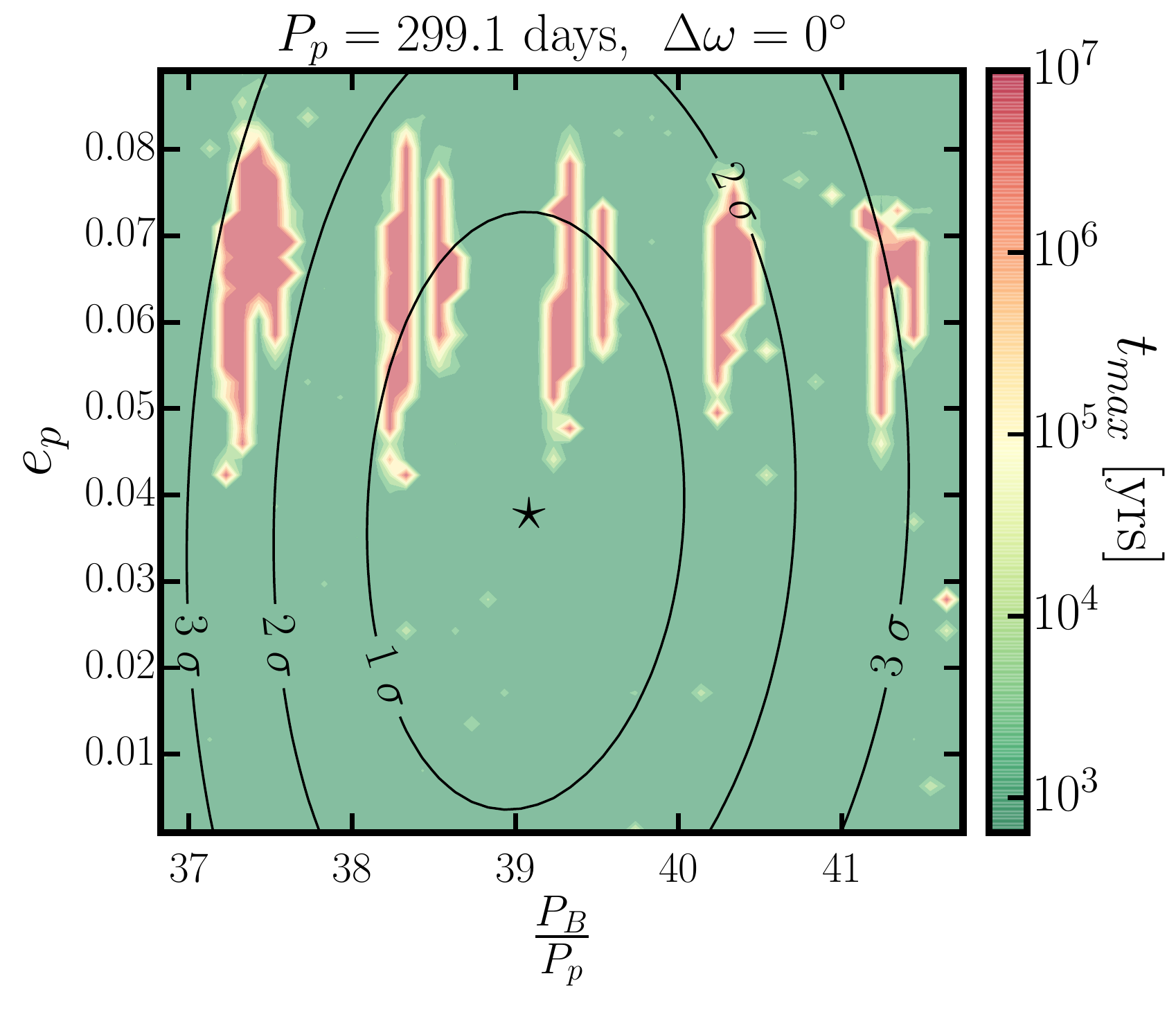}\\
\end{array} $

\end{center}

\caption{
Edge-on coplanar prograde grid in the  $P_B/P_p$ - $e_p$ space, where 
$P_p$ and $\omega_B$ are fixed at their best fit values of 299.1 days and 149.4$^\circ$, respectively, while
$\omega_p$ is also fixed to 149.4$^\circ$ to assure $\Delta\omega$ = 0$^\circ$. 
Color-coded is the time for which the system is stable. Red contours are the confidence levels corresponding to the
1$\sigma$, 2$\sigma$ and 3$\sigma$ confidence regions of the $\chi_{\nu}^2$ minimum (black star).
For this aligned grid, stability is achieved only in the range of initial 
$e_p$ $\approx$ 0.04 to 0.07 and $P_B$/$P_p \neq \mathbb{Z}$.
}

\label{p1p2_mvs_free}
\end{figure}

\begin{figure*}[btp]
\begin{center}$
\begin{array}{ccc} 
\includegraphics[width=18cm]{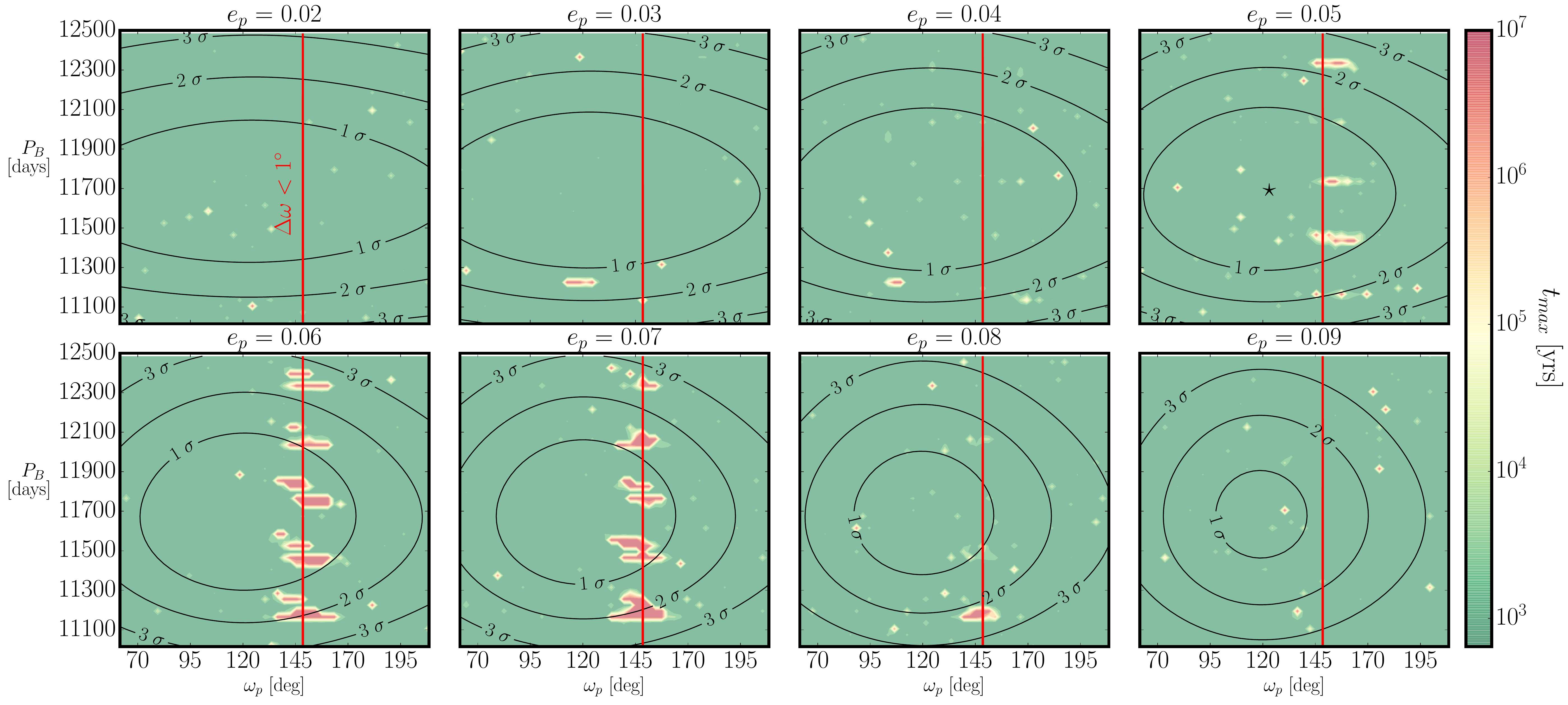}\\

\end{array} $

\end{center}

\caption{
Edge-on coplanar prograde grids for systematically varied $P_B$, $\omega_p$ and $e_p$ around 
 their best-fit values. Color-coded is the survival time, and black contours correspond 
 to the confidence levels of the best fit. 
Most of the fits are highly unstable, except those located in between an integer period ratio
 and near alignment ($\Delta\omega$ $\approx$ 0$^\circ$), while $e_p$ is between 0.05 and 0.07. 
Some of these stable islands are within the $1\sigma$ confidence level of the best fit and provide valid possibilities for the 
 orbital configuration of the HD~59686 system.
}

\label{p1p2_mvs}
\end{figure*}

\subsection{Coplanar prograde grids}
\label{prograde} 

Our edge-on prograde bootstrap analysis reveals that stable fits are clustered around initial
$\omega_p \approx 145^\circ$, $e_p$ = 0.05 to 0.07, and $P_B$/$P_p$ $\neq$ 39 or 40
(within 1$\sigma$ from the best fit). 
These three orbital parameters are also the least constrained parameters from our fitting, especially when 
compared to $P_p$, $\omega_B$ and $e_B$.
Therefore, any grid combination including $P_B$, $\omega_p$ and $e_p$ yields an adequate search for prograde coplanar stable fits. 
Figure~\ref{p1p2_mvs_free} shows stability results for an edge-on coplanar grid in the $P_B$/$P_p$ - $e_p$ space, where 
we fix $P_p$ and $\omega_B$ at their best-fit values of 299.1 days and 149.4$^\circ$, respectively.
Since from the bootstrap analysis we know that stability appears when $\Delta\omega$ $\approx$ 0$^\circ$ 
we keep $\omega_p$ = $\omega_B$ fixed, leading to an initially aligned configuration with $\Delta\omega$ = 0$^\circ$. 
In Figure~\ref{p1p2_mvs_free}, stability within the 1$\sigma$ confidence region from the best fit is achieved when $e_p$ $\approx$ 0.04 to 0.07 and
$P_B$/$P_p$ $\neq$ 39 or 40, which confirms the results from the bootstrap analysis.

As a next step, we test for stability in  eight $P_B$ - $\omega_p$ grids,
each with $e_p$ = 0.02 to 0.09 in steps of 0.01.
$P_B$ and $\omega_p$ are varied around the best coplanar fit in the range of 11000 to 12500 days and 60$^\circ$ to 210$^\circ$, respectively,
while the remaining parameters in the dynamical model are adjusted.
Figure~\ref{p1p2_mvs} shows the survival time resulting from this test together with the confidence levels.
Since these grids are constructed with three systematically varied parameters ($P_B$, $\omega_p$ and $e_p$),
we consider Figure~\ref{p1p2_mvs} as a three-dimensional parameter cube, where each grid is a separate $P_B$ - $\omega_p$ slice
 placed on a lower-resolution ``z''-axis constructed for different $e_p$. 
Thus, the significance levels (black contours) shown in Figure~\ref{p1p2_mvs} are calculated for three degrees of freedom. 
Clearly, most of the fits are highly unstable, except those located between integer period ratios, 
with nearly aligned orbits ($|\Delta\omega| \la 10^\circ$) and $e_p$ between 0.05 and 0.07.
We find that these stable islands cover the largest area when $e_p$ = 0.06.

\begin{figure*}[btp]
\begin{center}$
\begin{array}{ccc} 
\includegraphics[width=18cm]{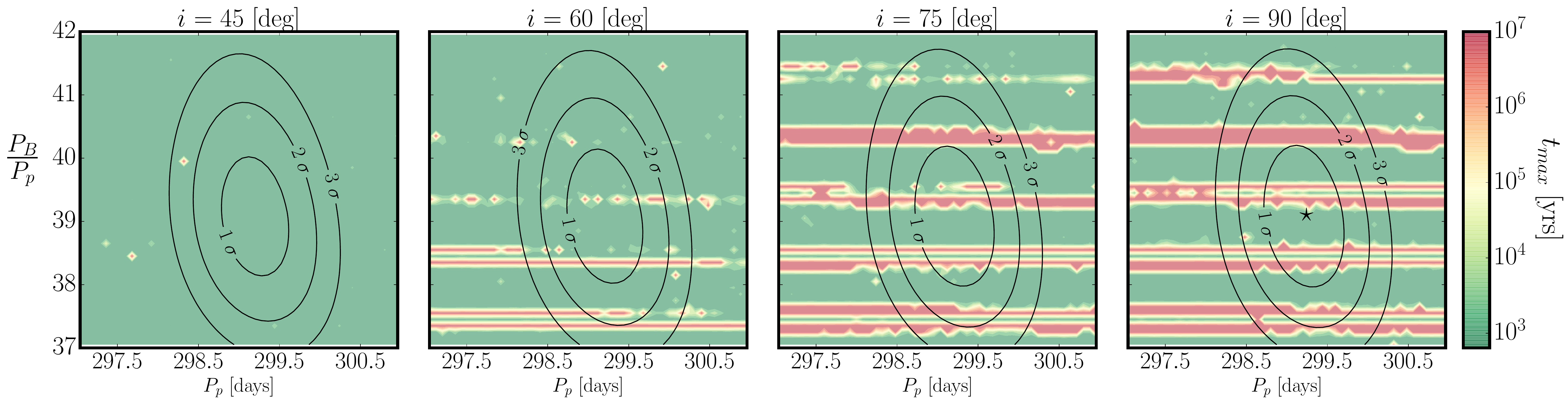}\\

\end{array} $

\end{center}

\caption{
 Coplanar prograde  grids of $P_B$/$P_p$ vs.\ $P_p$ for fixed $e_p$ = 0.06, 
 $\omega_p = \omega_B = 149.4^\circ$ (i.e., $\Delta\omega = 0^\circ$) and 
 $i$ = 90$^\circ$, 75$^\circ$, 60$^\circ$, 45$^\circ$, and 30$^\circ$.  
The stable regions seen in Figure~\ref{p1p2_mvs_free} are now extended through the whole 
range of $P_p$ and crosses inside the $1\sigma$ confidence levels of the best fit.
These stable regions exist down to $i$ = 60$^\circ$, below which both companions have
 masses large enough to make the system unstable.
}

\label{p1p2_mvs2}
\end{figure*}

Knowing that $e_p$ and $\omega_p$ are critical stability parameters, in Figure~\ref{p1p2_mvs2} 
we show results for $P_B/P_p$ - $P_p$ grids, where we fix 
$e_p$ = 0.06 and $\omega_p$ = $\omega_B$ = 149.4$^\circ$, and we systematically adopt 
inclination of $i$ = 90$^\circ$, 75$^\circ$, 60$^\circ$, 45$^\circ$ and 30$^\circ$.
In this way we study the stability of fits in the $P_B/P_p$ - $P_p$ parameter space 
for initially aligned orbits and increasing companion masses.
In these grids, the stable regions now extend through a large 
range of $P_p$ and cross inside the $1\sigma$ confidence regions.
Configurations with period ratios near an integer initially are highly unstable,
while stable configurations can be found between initial integer period ratios.
The stable regions are evident in the $i$ = 90$^\circ$ to 60$^\circ$ grids, but 
they become smaller with decreasing $i$ or increasing masses. 
The $i$ = 45$^\circ$ grid contains only a few marginally stable fits, which are likely unstable beyond 10 Myr,
while all configurations for the $i$ = 30$^\circ$ grid (not shown in Figure~\ref{p1p2_mvs2}) are unstable.
These results identify a lower limit for the inclination of $i \approx$ 45$^\circ$ for stable prograde coplanar configurations.
This inclination limit happens to coincide with the secondary star mass constraints discussed in \citet{Ortiz2016}.

\begin{figure*}[btp]
\begin{center}$
\begin{array}{ccc}

\includegraphics[width=8cm, height=4.6cm]{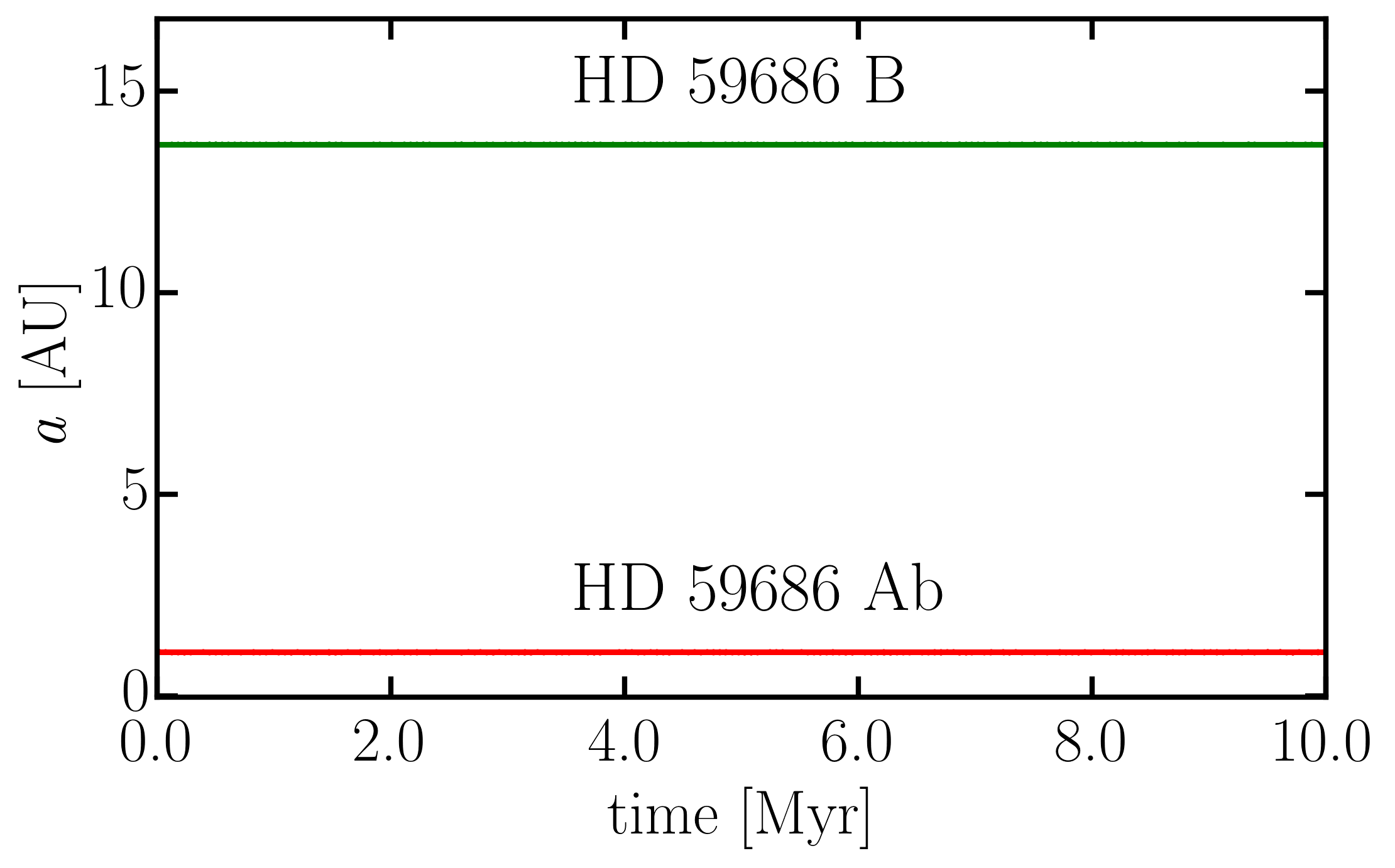}
\includegraphics[width=8cm, height=4.6cm]{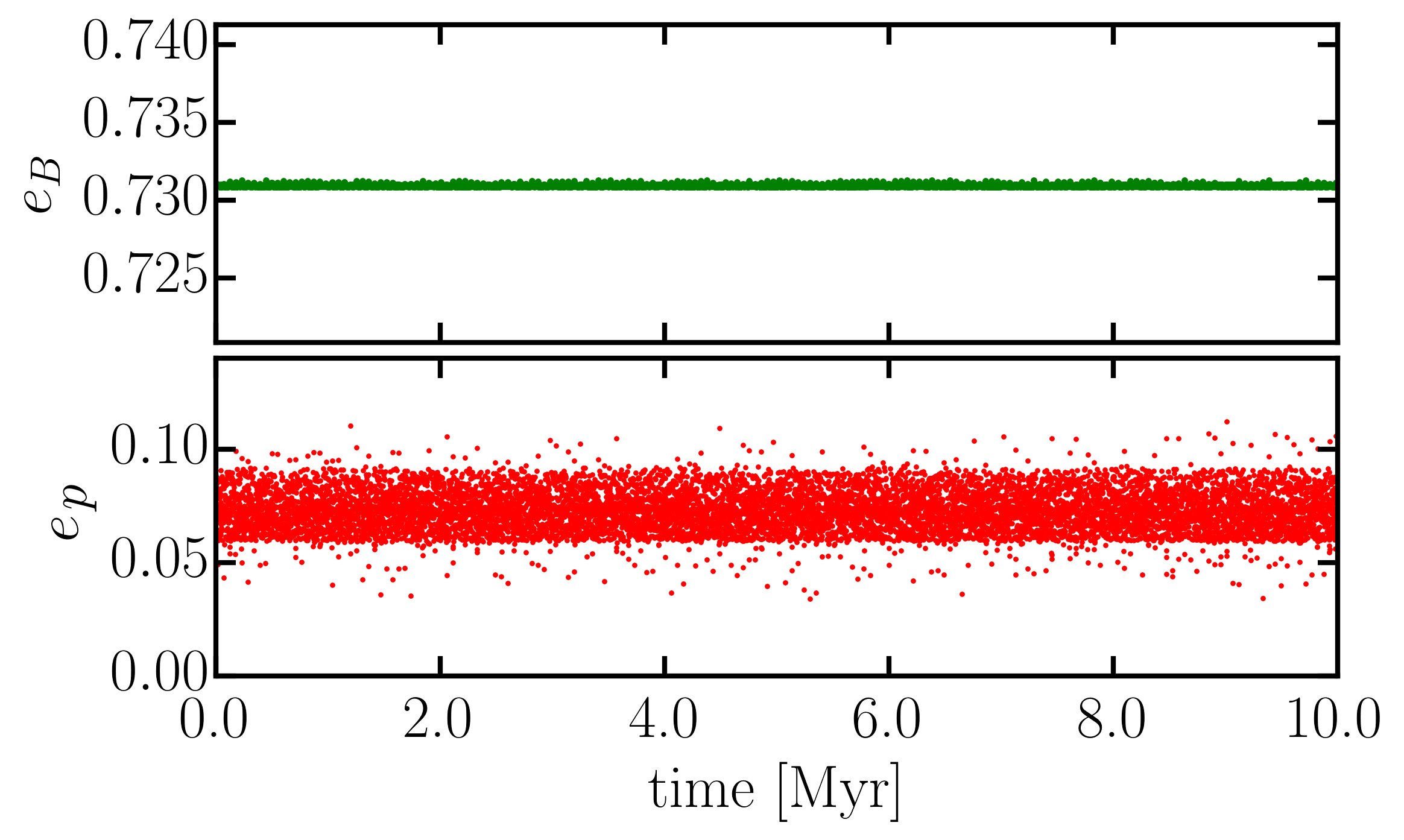} \\
\includegraphics[width=8cm, height=4.6cm]{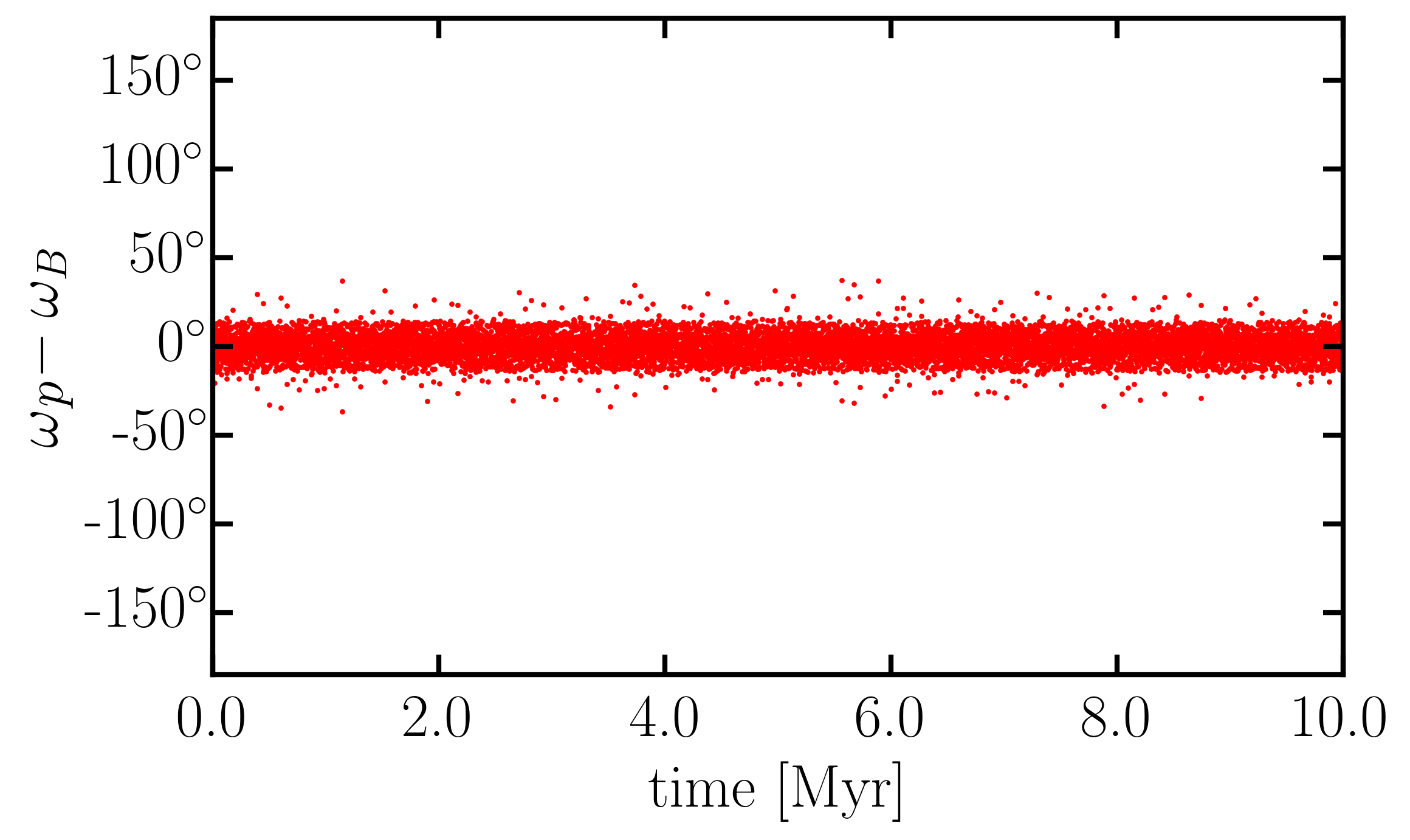}
\includegraphics[width=8cm, height=4.6cm]{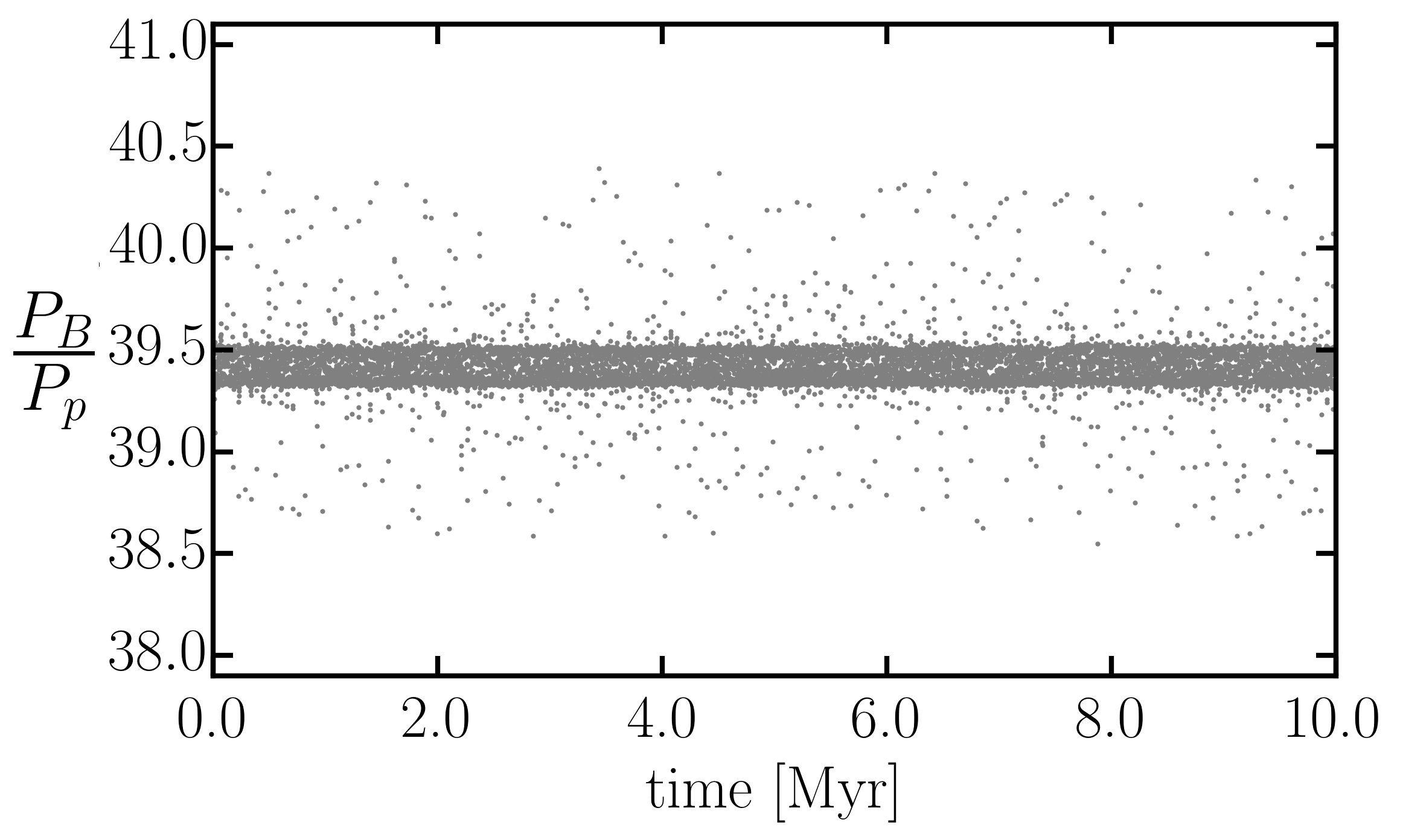}

\end{array} $

\end{center}

\caption{Top panels: Evolution of the binary (green) and planetary  (red) semi-major axes and eccentricities of the
best stable coplanar, edge-on and prograde fit with initial $P_B$/$P_p$ $\approx$ 39.3. 
No notable changes can be seen in $a_p$ and $a_B$.
The  binary eccentricity $e_B$ fluctuates with very small amplitude around 0.73, while $e_p$ librates with 
a larger amplitude between 0.04 and 0.11.
 Bottom panels: This fit is clearly locked in secular apsidal alignment, where the 
secular apsidal angle $\Delta\omega = \omega_p - \omega_B$ librates around 0$^\circ$ with a semi-amplitude of $\pm$37$^\circ$,
while the mean orbital period ratio is $P_B$/$P_p$ $\approx$ 39.4.
}

\label{stable_prograde}
\end{figure*}

Table~\ref{table:stable_fit} gives the orbital parameters and corresponding errors for the best stable fit
among these grids, which has an initial $P_B$/$P_p$ $\approx$ 39.3.
Figure~\ref{stable_prograde} shows the orbital evolution of this stable fit.
The evolution of the semi-major axes $a_p$ and $a_B$ in the upper left panel shows that 
this configuration remains long-term stable with well separated orbits. 
The binary eccentricity $e_B$ has very small amplitude variations around 0.73,
while the planet remains nearly circular, with $e_p$ varying between 0.04 and 0.11 (upper right panel). 
 The bottom two panels of Fig.~\ref{stable_prograde} show the
evolution of the secular  apsidal angle $\Delta\omega$ = $\omega_{\rm p} - \omega_{\rm B}$, which 
exhibits a clear libration around 0$^\circ$ with a semi-amplitude of $\pm$ 37$^\circ$,
while the mean period ratio during the integration is $\approx$ 39.4, close to the initial $P_B$/$P_p$.
We examine this configuration for librating resonance angles (see Eq.~\ref{eq:theta})
associated with the nearest 39:1 and 40:1 mean-motion commensurabilities,
and confirm that this stable prograde fit is not involved in a MMR.
 Such an orbital evolution is characteristic for secular apsidal alignment \citep[e.g.][]{LeeM2003,Michtchenko2004}. 
 The libration of $\Delta\omega$ around 0$^\circ$ is critical for the stability of our system, since it
 helps the lower-mass S-type object to retain small eccentricities while being significantly perturbed by the secondary star. 
We have investigated all stable prograde islands shown in Figures~\ref{p1p2_mvs_free}, \ref{p1p2_mvs} and \ref{p1p2_mvs2}, 
and they all exhibit similar evolution, with librating
$\Delta\omega$ around 0$^\circ$, circulating MMR angles $\theta_{1,n}$, small $e_p$ and a non-integer mean period ratio $P_B$/$P_p$.
Thus, we conclude  that if the HD~59686 system is indeed prograde, then it
must be locked in secular apsidal alignment to stabilize the orbits.

Two additional remarks on the secular apsidal alignment and non-MMR nature of the 
stable islands in Figures \ref{p1p2_mvs_free},~\ref{p1p2_mvs} and \ref{p1p2_mvs2} are in order. First, one may be concerned that the 
non-integer initial and mean period ratios may not represent the true period ratio due 
to the large mass of the secondary star. However, the Hamiltonian in Jacobi coordinates 
in Equation (11) of \cite{LeeM2003} shows that the perturbations to the Keplerian motions 
from the interactions between the secondary star and the planet remain small and the semimajor 
axes and period ratio should be nearly constant, throughout most of the binary orbit 
(including the initial epoch when the secondary is $\sim 20\,$AU from the primary). 
Even when the secondary is at periastron, we can estimate from the lowest order term 
in the perturbations to the Kepler motions in Equation (11) of \cite{LeeM2003} that 
the full amplitude of the variation in the period ratio should be $\sim (9/2) (m_B/M_{\ast}) (a_p/a_B (1-e_B))^3 \sim 3.3\%$ 
if $e_p$ is small. These results are consistent with the evolution of $P_B/P_p$ shown in Figure~\ref{stable_prograde},
where $P_B/P_p$ is near the initial value most of the time and shows scatter of $\sim 4.5\%$.

Second, Wong \& Lee (in preparation) have systematically studied the stability of 
circumprimary planetary orbits in the HD~59686 system, with initial conditions in 
grids of $a_p$ and $e_p$ for several values of $\Delta\omega$ and mean anomalies. 
For the coplanar prograde case, they confirm the existence of islands stabilized 
by secular apsidal alignment. They also find islands that are stabilized by MMR, 
but these are at higher $e_p$ and do not fit the observed planet.

\subsection{Coplanar retrograde grids}
\label{retrograde}

We repeat the grid analysis for retrograde coplanar configurations by fixing $\Delta i$ = 180$^\circ$ for each fit.
We find that all fits within 3$\sigma$ from the best fit are stable for at least 10 Myr. 
Particularly, all fits in the $P_B/P_p$ - $P_p$ grids are stable despite the large 
companion masses for $i$ = 30$^\circ$ and even $i$ = 15$^\circ$. 
The stability of the $P_B/P_p$ - $P_p$ grids in the range $i$ = 90$^\circ$ to 15$^\circ$
is in agreement with the results presented in Figure~\ref{incl_reduced_chi_retro} and Section \S\ref{Inclined coplanar},
where all the retrograde coplanar inclined best fits are stable down to $i$ = 5$^\circ$.
We conclude that the best retrograde coplanar fit is well within a large stable phase space region,
not necessary involved in a MMR.
Therefore, no meaningful stability constraints can be obtained from the retrograde coplanar grids,
except that the retrograde orbits yield very strong candidates for the HD~59686 system configuration.

\subsection{Mutually inclined grids}
\label{mutually}

\begin{figure*}

\begin{center}$
\begin{array}{ccc} 

\includegraphics[width=18cm]{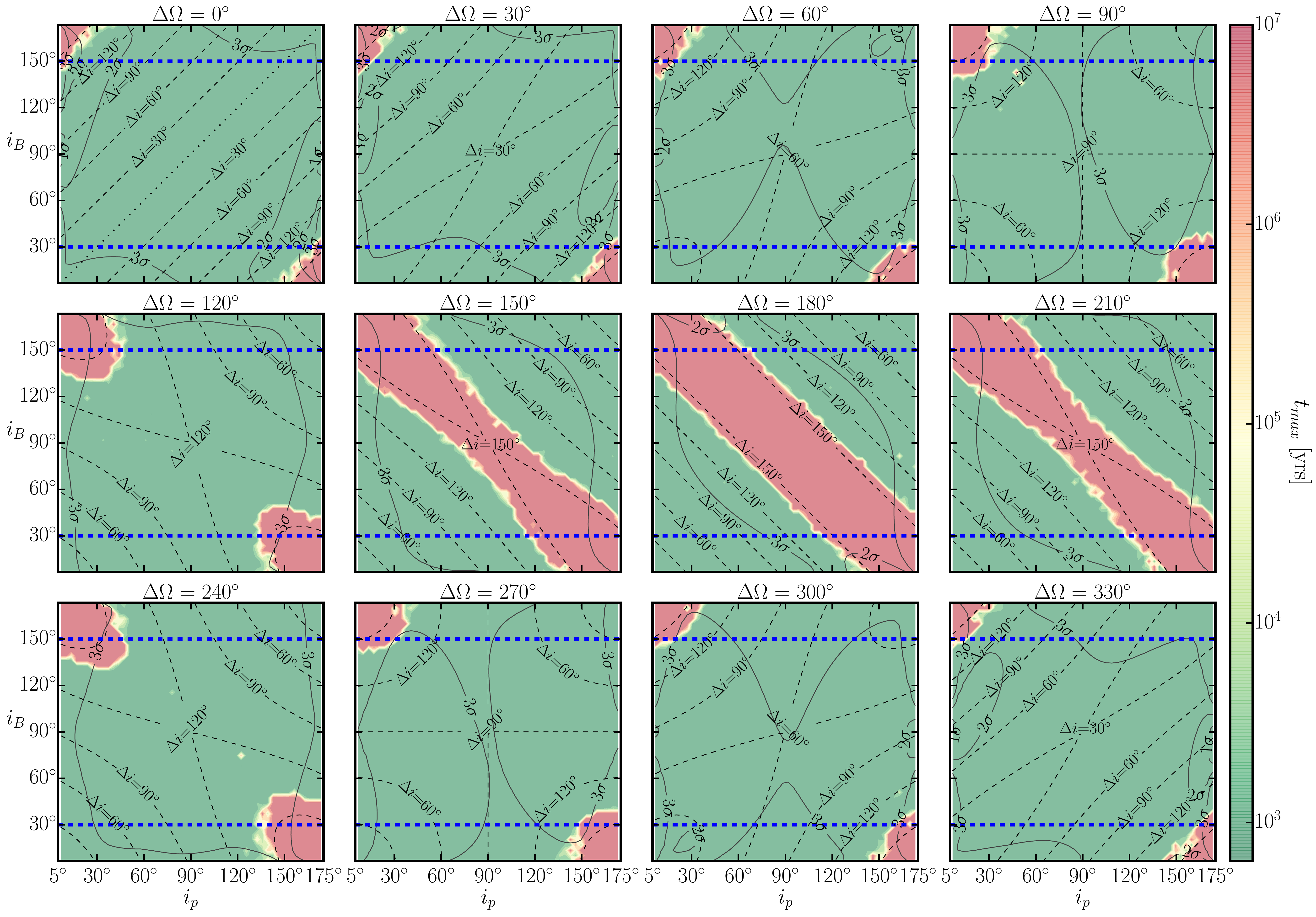}  \\

\end{array} $

\end{center}

\caption{ Mutually inclined grids for different $\Delta\Omega$ covering (almost) all possible mutual inclined 
configurations. 
The black dashed contours are the initial mutual inclination borders with steps of $\Delta i$ = 30$^\circ$. 
The stability is tested for 10 M\,yr, and red filled contours show the grid areas where the orbits are stable. 
The HD~59686 S-type companion is stable for a large set of dynamical masses
only when it is on a nearly coplanar retrograde orbit with 145$^\circ$~$\lesssim$~$\Delta i$~$\leq$~180$^\circ$
with respect to the binary plane. }
\label{i1i2} 
\end{figure*}

We construct twelve separate $i_B$ - $i_p$ grids, where for each grid we adopt $\Delta\Omega =
0^\circ$ to 330$^\circ$, with a step size of 30$^\circ$.
In these grids $i_B$ and $i_p$ are varied from 5$^\circ$ to 175$^\circ$, with a step size of $3.4^\circ$,
corresponding to 50 different values.
Fits with $i_p$, $i_{B}$ below 5$^\circ$ and above 175$^\circ$ are not considered, since these inclinations 
lead to nearly face-on orbits and the dynamical masses of the companions will be very large.
In fact, as discussed in Section~\S\ref{Inclined coplanar}, other observations constrain 
$\sin i_{B}$ to values greater than 0.5, 
but since the dynamical model does not reject more massive secondary stellar companion, 
we construct symmetrical grids with the same range of $i_p$ and $i_{B}$.
Thus, we study mutually inclined configurations covering almost all possible system geometries.
The results from this test are shown in Fig.~\ref{i1i2}.

\begin{table}[tb]
\begin{adjustwidth}{-5cm}{} 

\resizebox{0.8\textheight}{!}
{\begin{minipage}{1.0\textwidth}

\centering   

\caption{Stable Prograde Fit with $P_B$/$P_p$ $\approx$ 39.3}   

\label{table:stable_fit}      

\begin{tabular}{ lccccc}     

\hline\hline  \noalign{\vskip 0.7mm}      


  

Parameter & ~~~HD~59686 Ab~~~ & ~~~HD~59686 B~~~  \\

\hline\noalign{\vskip 0.5mm}

$K$  [m\,s$^{-1}$]                        & 137.0 $\pm$ 3.3     &  4011.6 $\pm$ 3.7         \\
$P$ [days]   			          & 299.2\tablenotemark{a}       &  11772.7\tablenotemark{a}      \\  
$e$                                       & ~~~~0.06\tablenotemark{a}   &  ~~~~0.731  $\pm$ 0.005   \\
$\omega$ [deg]                            & ~~149.3 \tablenotemark{a}  &  149.4   $\pm$ 0.1         \\   
$M_0$ [deg]                               & ~~272.0 $\pm$ 1.4  &  259.8   $\pm$ 0.4         \\  \noalign{\vskip 0.9mm} 
RV$_{ \mathrm{off}}$~[m\,s$^{-1}$]        & ~~243.2 $\pm$ 4.0  &                            \\ \noalign{\vskip 0.9mm} 
$i$ [deg]                                 & 90.0\tablenotemark{a}       & 90.0\tablenotemark{a}              \\ 
$\Omega$ [deg]                            & 0.0\tablenotemark{a}        & 0.0\tablenotemark{a}               \\ 
$\Delta i$ [deg]                          & 0.0                 &                            \\ \noalign{\vskip 0.9mm} 
$a$ [AU]                                  & 1.09                &  13.67                     \\  
$m \sin i$ [$M_{\mathrm{Jup}}$]           & 6.97                &  558.5                    \\ \noalign{\vskip 0.9mm}
$r.m.s. $ [m\,s$^{-1}$]                   & 19.84               &                            \\ 
$\chi^2$                                  & 78.57               &                            \\
$\chi_{\nu}^2$                             & 0.970               &                            \\

\noalign{\vskip 0.5mm}

\hline\hline\noalign{\vskip 1.2mm}

\end{tabular}  

\end{minipage}}
 
\end{adjustwidth}
\tablenotetext{a}{\small Fixed parameters.}

\end{table}

The only stable region we find in these grids corresponds to nearly coplanar retrograde geometries with $\Delta i >$ 145$^\circ$.
The orbital evolution of these stable fits is very similar to that
of the best retrograde coplanar fit discussed in \S\ref{Prograde and retrograde} and shown in the middle panel of Fig.~\ref{best_fits_evol}.
For nearly coplanar retrograde orbits, the amplitude of the variations in the inner companion eccentricity $e_p$ is rather large. 
It is about 0.35 for strictly coplanar orbits and increases with increasing mass 
of the secondary star (i.e., with decreasing $\sin i_B$) to about 0.42 at maximum, when $i_B$ and $\Delta i$ $\approx$ 145$^\circ$. 
Meanwhile, the $e_p$ and $e_B$ oscillation frequency is highest in the coplanar configuration and decreases with decreasing $\Delta i$.
For fits near the stability boundary of $\Delta i$ $\approx$ 145$^\circ$, another shorter term eccentricity variation is visible on top of the main secular 
eccentricity oscillation, which has a period of a few hundred years.
In this stable region, $\Delta i$ also exhibits small variations around the initial best fit value, except for the coplanar case, where $\Delta i$ remains constant at 180$^\circ$.
 We do not find evidence that any of the stable retrograde fits are in MMR.

The small $\sin i_p$ retrograde corners of Fig.~\ref{i1i2} are 
intriguingly stable, suggesting that the inner companion's 
 dynamical mass could be as high as 10 times its minimum mass
(i.e.\ $\sin i_p \approx$ 0.1), converting the planet to a massive brown dwarf 
or at the extreme even a very low-mass M-dwarf star.
A massive, highly inclined and retrograde inner companion may be consistent with
the Hipparcos astrometry and dynamical modeling, but 
to preserve the system's stability the outer binary companion must also have a small $\sin i_B$ and that is  not supported by observations.
On the one hand, if we assume that the outer companion is a main sequence star, then we can limit its
inclination to $150^\circ \ge i_B \ge 30^\circ$ (marked by a blue dashed lines in Fig.~\ref{i1i2}), beyond which it would be a solar mass
star and should have been detected in \cite{Ortiz2016}. 
All stable solutions with secondary $150^\circ \ge i_B \ge 30^\circ$ support a planet mass 
object for the inner companion.
On the other hand, the LBT observations obtained in \cite{Ortiz2016} would not
be sensitive to a white dwarf secondary with $\sim$ 1 $M_\odot$.
Thus, in principle, both companions of HD~59686 can have very small $\sin i$, making the system a
hierarchical retrograde triple of a K-giant, M-dwarf and a white dwarf.
Such an exotic system would be stable in a retrograde orbit, but then
in all cases the excited inner-binary eccentricity would 
oscillate with much larger values than the one currently observed, i.e., our observations would have caught the system at a very special time.
Therefore, we conclude that the inner companion is most likely of planetary origin.

 Overall, we identify a large and confident stable region for the HD~59686 system, which turns out to be at high mutual inclinations 
of $\Delta i$~$\gtrsim$~145$^\circ$, corresponding to nearly coplanar, but retrograde orbits. 
Mutual inclinations with $\Delta i$ between 30$^\circ$ and 145$^\circ$
lead to instability on very short time scales due to Lidov-Kozai effects,
and thus such orbital configurations are very unlikely.
Nearly coplanar and prograde best fits with $\Delta i \la 30^\circ$ are also unstable.

\section{Summary and Conclusions}
\label{Discussion}

HD~59686 is without any doubt a very interesting three-body system. 
It consists of a single-lined spectroscopic binary with a K~giant primary
with $M$ = 1.92~$M_{\odot}$, a low-mass secondary star with 
at least $m_B$~=~0.53~$M_{\odot}$ and at most $\sim$ 1 $M_{\odot}$, 
and an additional S-type planet with at least 7~$M_{\mathrm{Jup}}$. 
This system has a challenging architecture, since the secondary star orbits beyond the planet orbit ($a_p$ = 1.09 AU) 
on a relatively close ($a_B$ = 13.6 AU) and very eccentric ($e_B$ = 0.73) orbit.
As a result, the binary periastron distance is only $q_B$ = 3.7~AU, leading to strong
interactions with the planet, and thus challenging the system's long-term stability.

In this paper, we performed a detailed orbital and stability analysis of the HD~59686 system via dynamical modeling of RV data and
long-term $N$-body integrations.
We aimed to refine the orbital parameters by stability constraints, which can provide clues on the formation history.
This is important since only a handful of S-type planetary candidates in compact binary systems are
known in the literature, and the HD~59686 system illustrates how planets could form and remain stable in an S-type orbit 
around a star under the strong gravitational influence from a close stellar secondary.

Our global best fit with the lowest $\chi_{\nu}^2$ suggests a triple star system with nearly polar 
orbits ($\Delta i$ = 92.7 $\pm$ 3.3$^\circ$), instead of a binary with an S-type planet. 
We have shown, however, that such orbits quickly lead to instability due to the Lidov-Kozai effect.  
Within only 600 yrs the Lidov-Kozai effect 
excites the eccentricity of the inner object to a value which leads to collision with the primary K~giant star.
Orbital fits with parameters similar to the near-polar global best fit are very unstable
and experience a similar fate, although they all have lower $\chi_{\nu}^2$ values compared to the 
fits corresponding to a coplanar configuration.
We conclude that the near polar configurations cannot represent the true system configuration, 
and that their small $\chi_{\nu}^2$ values are most likely a result of model degeneracies, 
which come from the limited number and accuracy of RV data points. 
These conclusions are supported by our bootstrap statistical analysis.

We find that HD~59686's planet can survive only on nearly coplanar, most likely retrograde orbits.    
We find that when the system's mutual inclination is $\Delta i$ = 180$^\circ$ 
(i.e., coplanar and retrograde), the system is fully stable for 
a large set of orbital solutions and companion masses. 
Long-term stability is also preserved for nearly coplanar retrograde configurations 
with 145$^\circ$~$\lesssim$~$\Delta i$~$\leq$~180$^\circ$.

Although most of the coplanar prograde fits consistent with HD~59686's RV data are unstable, 
we have demonstrated that stable prograde fits in fact do exist.   
Our bootstrap and grid search analysis shows that a fraction of prograde fits  
(mostly within $1\sigma$ from the best fit) are stable for at least 10~Myr. 
These fits are located in narrow strips of the orbital period space where the initial period ratio $P_B$/$P_p$ 
is not an integer number, with the best chances of stability near $P_B$/$P_p$ $\approx$ 38.4 and 39.4.
Therefore, the system can survive only between high-order MMR, while the MMR themselves 
 have a destabilizing effect on the S-type planet.
However, we find that the planetary $e_p$ and $\omega_p$ are also very important parameters 
which control the planet stability. 
The stability results indicate that the planet must be initially on a nearly aligned orbit 
with the binary ($\Delta\omega \leq 10^\circ$) and have $e_p$ $\approx$ 0.06.
In such a configuration, long-term stability is ensured by secular apsidal alignment
between the binary and planetary orbits, with $\Delta\omega$ 
librating around $0^\circ$ with relatively small amplitude.

The stable islands shrink when we assume lower inclinations and more massive companions, while keeping coplanarity. 
Below $i = 45^\circ$, all prograde coplanar configurations are unstable,
which suggests that if the system in indeed prograde and coplanar, then
90$^\circ$ $\geq$ $i$ $\geq$ 45$^\circ$, with most stable fits at $i$ = 90$^\circ$.
The orbital dynamics of these stable prograde fits with larger masses similarly
shows secular apsidal alignment where  $\Delta\omega$ librates around 0$^\circ$.

As a final discussion point, we note that there are arguments in favor of and against
both prograde and retrograde configurations. 
For example, the retrograde stable region is very large and it can explain the RV data
with great confidence, but forming a retrograde planet requires some exotic 
scenarios \citep[see discussion in ][]{Ortiz2016}. 
On top of that, looking at the retrograde planet eccentricity evolution (middle panel of Figure~\ref{best_fits_evol}), we estimate that
$\sim$ 18\% of the time $e_p < 0.1$ (within 3$\sigma$ of the best-fit value),
and only $\sim$ 8\% of the time $0.03 < e_p < 0.07$ (within 1$\sigma$).
If the system is indeed in a retrograde configuration, then 
the Lick RVs must have been obtained in a phase which has rather low probability when $e_p$ is as low as $\sim 0.05$.
On the other hand, small $e_p$ is not a problem for the prograde stable fits where 
most of the time the planet eccentricity is in the range $0.04 < e_p < 0.11$ (Figure~\ref{stable_prograde}). 
However, the stable prograde islands are very narrow, and whether it is possible to form a massive planet in a narrow stable region
in secular apsidal alignment with the eccentric close binary is a problem that deserves 
a closer look in the future.

\acknowledgments

T.T and M.H.L. are supported in part by Hong Kong RGC grant HKU 17305015.
We thank the anonymous referee for the excellent comments that helped to improve this paper. 
      
\bibliographystyle{apj}

\bibliography{Trifonov_2015}

\end{document}